\documentclass[12pt,english]{article}
\usepackage[T1]{fontenc}
\usepackage[latin9]{inputenc}
\usepackage[dvipsnames]{xcolor}
\usepackage{mathtools}
\usepackage{amsmath}
\usepackage{amssymb}
\usepackage{bbold}
\usepackage{stmaryrd}
\usepackage{graphicx}
\usepackage{dsfont}
\usepackage{soul}

\makeatletter
\usepackage{jheppub}
\def\@fpheader{\relax}
\allowdisplaybreaks[1]

\makeatother

\usepackage{babel}
\begin{document}


\title{Canonical Electrodynamics in Ashtekar-Barbero variables}

\author[a]{Federica Fragomeno,}
\author[a,b,c]{Saeed Rastgoo,}
\affiliation[a]{Department of Physics, University of Alberta, Edmonton, Alberta T6G 2G1, Canada} 
\affiliation[b]{Department of Mathematical and Statistical Sciences, University of Alberta, Edmonton, Alberta T6G 2G1, Canada}
\affiliation[c]{Theoretical Physics Institute, University of Alberta, Edmonton, Alberta T6G 2G1, Canada}
\emailAdd{ffragome@ualberta.ca}
\emailAdd{srastgoo@ualberta.ca}

\abstract{
We present the full Hamiltonian analysis of electrodynamics, including fermions, photons, and their interaction, on a general curved spacetime in Ashtekar-Barbero variables. Only the time gauge is imposed and the theory is background independent. The details of the constraints of the theory for fermionic and electromagnetic fields coupled to gravity are presented. Moreover, the equations of motion are derived and it is shown that they reduce to the Dirac equation, Maxwell equation and their interaction term in curved or flat spacetime in the appropriate limits. Finally, the behavior of the theory under parity transformation is analyzed. This result presents a suitable foundation for loop or polymer quantization.
}
\maketitle

\section{Introduction}

Fermions comprise the realistic matter in the universe. Understanding many astrophysical phenomena, particularly those that are relevant for tests of gravity in strong or quantum regimes, need a careful and detailed study of fermions, particularly in the electromagnetic sector, in those regimes. Therefore, it is of utmost importance to have a general background independent formulation of fermions, i.e., one that includes all the backreactions. 

Having a classical canonical theory in hand which describes the interaction of these three ingredients is a first and crucial step in being able to proceed with canonical quantization in approaches such as loop quantum gravity (LQG) and generalized uncertainty principle (GUP). Furthermore, there are several interesting symmetry reduced scenarios, such as dynamical (shell collapse, Oppenheimer-Snyder scenario) black hole solutions in the effective regime of LQG, where only scalar fields have been incorporated. Since these models are almost all canonical systems, a readily Hamiltonian formulation where fermions and photons are also involved will open new possibilities for extending these effective models to more realistic scenarios.



There are previous works done with regard to fermions in a canonical theory in Ashtekar-Barbero variables \cite{Bojowald:2010qpa, Das:2008r,Thiemann:1998QSDV, Jacobson:1988qta, Mercuri:2006wb, PhysRevD.73.084016, Date:2011rd, PhysRevD.103.124030, Duque:2025spv, Romero:2022kjh} that paved the way for our work. However, they usually ignore the interaction term between the fermions and the photons and hence do not present the full canonical electrodynamics (CED) Hamiltonian in a non-fixed general curved background. Some of them also lack the details of computation or seemingly suffer from some missing terms or some inconsistencies. In particular, it seems that in \cite{ Das:2008r} the choice of the definition of the spin connection $\Gamma_{a}^{i}$ is not compatible with their covariant derivative $D_a=\partial_a+A_a^i \tau_i$ and their constraints. In \cite{Thiemann:1998QSDV}, the authors present the Hamiltonian constraint for gravity plus fermions, however, it seems that they do not start directly from the action, and as a consequence there is a missing term in the modification of the Ashtekar connection as well as in the fermionic Hamiltonian constraint. They also seem to take the modification of the Ashtekar connection derived in \cite{Jacobson:1988qta}. However, the latter starts from the Palatini action rather than the Holst action
. As a consequence it does not seem to involve axial fermionic current and only incoporates fermionic current. It also seems that the fact that the extrinsic curvature in presence of fermions is no longer symmetric
because of the torsion contribution is overlooked. As a consequence, in the scalar constraint, a term proportional to $\epsilon_i{ }^{jk} K_a^i E_j^a J_k $ is missing. 

In this work we set to fill the aforementioned gaps. We carefully study a generic background-independent system of gravity, fermions and photons, starting from the action and making a $3+1$ decomposition, writing the theory in Ashtekar-Barbero variables. This is done for both Weyl and Dirac fermions, in a step by step, clear, and detailed manner. We then derive the modification to both the canonical variables, and the constraints, showing in a clear way the distinct contribution of each sector of to these constraints. We also provide the equations of motion of fermions in this setting and study the behavior of the system under parity transformation. Another difference of this work compared to the previous ones, is that we do not modify the gamma matrices, and rather change the identities between them to incorporate the changes needed for the background-independent framework. This way it is easier to go back to a flat field theory. Furthermore, as mentioned before, to our knowledge, this is the first time that the CED interaction term is studied in detail in such a canonical formulation in Ashtekar-Barbero variables. 

The structure of this paper is as follows: In Sec. \ref{vacuum}, we provide a brief overview of vacuum canonical gravity in Ashtekar-Barbero variables starting from the Holst action. Then in Sec. \ref{sec:fermions}, we present how fermions are incorporated in the theory, and how they modify the canonical variables and constraints. Using the results of the previous sections, in Sec. \ref{sec:QED} the full CED Hamiltonian in an arbitrary dynamical background is constructed by incorporating both the photon field and its interaction with fermions and gravity. In Sec. \ref{sec:parity} we study the behavior of the system under the parity transformations. Finally, in Sec. \ref{sec:Conclusion} we summarize and conclude the paper.

\section{Brief overview of Ashtekar-Barbero variables \label{vacuum}}
In this section we will quickly review Ashtekar-Barbero formalism
of GR in order to introduce the notation used in the Fermions and
CED Sections.

We start with a first order action of GR, the \textit{Holst action}
\cite{Bojowald:2010qpa} 
\begin{equation}
S=\frac{1}{2\kappa}\int d^{4}x|\text{det}\,e|e_{I}^{a}e_{J}^{b}P_{KL}^{IJ}F_{ab}^{KL}(\omega)\,,
\end{equation}
written in terms of the spacetime tetrads $e_{I}^{a}$ and their determinant
$(\text{det}\,e)^{-1}$. The (co)tetrads obey 
\begin{equation}
g_{ab}=\eta_{IJ}e_{a}^{I}e_{b}^{J}.
\end{equation}
Throughout this work the first four lower case Latin indices $a,\,b,\, c, \, d$ are abstract spacetime (sometimes spatial)
indices. The rest of the lower case Latin indices are $su(2)$ indices. Upper case Latin indices are the internal
Lorentz indices associated to the $SO(1,3)$ gauge freedom under which
the tetrads transform. The connection one form associated to this
transformation is $\omega_{a}^{IJ}$, and $F^{IJ}(\omega)=d\omega^{IJ}+\omega^{I}{}_{K}\wedge\omega^{KJ}$
is the antisymmetric 2-curvature of $\omega_{a}^{IJ}$. Furthermore
$\kappa=8\pi G$ and 
\begin{equation}
P_{KL}^{IJ}=\delta_{K}^{[I}\delta_{L}^{J]}-\frac{1}{2\beta}\epsilon^{IJ}{}_{KL},
\end{equation}
where $\beta$ is the \textit{Barbero-Immirzi parameter}. The term
in the action corresponding to the second term on the right hand side
of $P_{KL}^{IJ}$ is a topological term related to the Nieh-Yan invariant,
thus it will not affect the classical equation of motion in the vacuum case. However,
its presence is essential for obtaining the new variables that will
be use hereafter.

Varying the action with respect to the connection, $\tfrac{\delta S}{\delta\omega_{a}^{IJ}}=0$,
yields the \textit{compatibility condition} 
\begin{equation}
P^{KL}{}_{IJ}\mathcal{D}_{b}\left(|e|e_{K}^{[a}e_{L}^{b]}\right)=0\,.
\label{comp_cond_vac}
\end{equation}
where the covariant derivative is defined as $\mathcal{D}_{a}v^I = \partial_av^I + \omega_a^I\,_J v^J$, with the indices carrying their aforementioned meanings and $v^I$ being an $SU(2)$-vector.
This condition expresses the compatibility between the tetrad and the spin connection. This ensures that all tensor fields (such as the Dirac spinor field, the metric, etc.) transform properly under local Lorentz transformations.

To proceed toward a canonical formulation, one foliates the spacetime into spatial slices following the ADM decomposition. For this purpose, a spacetime tensor field, the triad $\mathcal{E}_{I}^{a}=e_{I}^{a}+n_{I}n^{a}$, is introduced where $n^{a}$ is the unit timelike vector field normal to the spatial slices, and $n_{I}\coloneqq e_{I}^{a}n_{a}$. It is also common to fix the boost part of the internal Lorentz transformations by imposing the ``time gauge'' condition $e_{0}^{a}=n^{a}$ \cite{Bojowald:2010qpa}.

We can now express the Holst action in terms of the components of
the unit normal vector
\begin{equation}
n^{a}=N^{-1}(t^{a}-N^{a})\label{eq:na-normal}
\end{equation}
where $N$ is
the \textit{lapse function}, $N^{a}$ the \textit{shift vector} and $t^a$ is the \textit{time-evolution vector field} defined such that $t^a\nabla_at=1$.
Making use of the relation $|\text{det}\,e|=N\sqrt{\text{det}\,h}$, where $\text{det}\,h$ is the determinant of the spatial metric $h_{ab}$ induced on the spatial hypersurfaces, the action
becomes 
\begin{equation}
S=\frac{1}{2\kappa}\int d^{4}x\sqrt{\text{det}\,h}P^{IJ}{}_{KL}F_{ab}^{KL}(\omega)(-2t^{a}n_{I}+N\mathcal{E}_{I}^{a}+2N^{a}n_{I})\mathcal{E}_{J}^{b}\,.
\end{equation}
As we will see below, the first term of the action leads to the symplectic
term from which the canonical variables can be read off, plus two
constraints (which both turn out to be the same, namely the Gauss
constraint). The second and third terms will yield the Hamiltonian
and the spatial diffeomorphism constraints, respectively. 

Using the definition of the Lie derivative given by $\mathcal{L}_{t}X_{b}^{i}=t^{a}\partial_{a}X_{b}^{i}+X_{a}^{i}\partial_{b}t^{a}$,
the first term of the above action can be further decomposed as 
\begin{equation}
\begin{aligned} & \frac{1}{\kappa\beta}\int d^{4}xE_{j}^{b}\mathcal{L}_{t}\left(\beta K_{b}^{j}+\Gamma_{b}^{j}\right)+\frac{1}{\kappa\beta}\int d^{4}x\Bigl\{\beta\omega_{t}^{0j}\left(\partial_{b}E_{j}^{b}+\epsilon_{jk}\,^{l}\Gamma_{b}^{k}E_{l}^{b}-\frac{1}{\beta}\epsilon_{jk}{}^{l}K_{b}^{k}E_{l}^{b}\right)\\
 & -\frac{1}{2}\epsilon_{kl}^{j}\omega_{t}^{kl}\left(\partial_{b}E_{j}^{b}+\beta\epsilon_{jm}{}^{n}K_{b}^{m}E_{n}^{b}+\epsilon_{jm}{}^{n}\Gamma_{b}^{m}E_{n}^{b}\right)\Bigr\}\,,
\end{aligned}
\label{eq:S-first-term}
\end{equation}
where we have introduce the \textit{densitized triad}, the \textit{extrinsic
curvature} and the \textit{spin connection} compatible with the triad,
respectively as 
\begin{align}
E_{i}^{a}= & \sqrt{\text{det}\,h}\,\mathcal{E}_{i}^{a}\,,\\
K_{a}^{i}= & \omega_{a}^{0i}\,,\\
\Gamma_{a}^{i}= & -\frac{1}{2}\epsilon^{i}{}_{jk}\omega_{a}^{jk}\,,
\end{align}
with the choice $\epsilon^{0ijk}=\epsilon_{tabc}=1$ for the Levi-Civita
symbol. It is now seen that the first term of \eqref{eq:S-first-term}
is the symplectic term of the action from which we can read off the
configuration variable, the \textit{Ashtekar-Barbero connection},
defined as 
\begin{equation}
A_{a}^{i}=\beta K_{a}^{i}+\Gamma_{a}^{i}\,,
\end{equation}
and its conjugate momentum the densitized triad $E_{i}^{a}$. Using
this definition and with further manipulation, Eq. \eqref{eq:S-first-term}
becomes 
\begin{equation}
\begin{aligned} & \frac{1}{\kappa\beta}\int d^{4}xE_{j}^{b}\mathcal{L}_{t}A_{b}^{j}+\frac{1}{\kappa\beta}\int d^{4}x\Bigl\{\beta\omega_{t}^{0j}\left(\mathcal{D}_{b}^{(A)}E_{j}^{b}-\frac{1+\beta^{2}}{\beta^{2}}\epsilon_{jk}{}^{l}K_{b}^{k}E_{l}^{b}\right)-\frac{1}{2}\epsilon_{kl}^{j}\omega_{t}^{kl}\mathcal{D}_{b}^{(A)}E_{j}^{b}\Bigr\}\\
 & =\frac{1}{\kappa\beta}\int d^{4}xE_{j}^{b}\mathcal{L}_{t}A_{b}^{j}+\frac{1}{\kappa\beta}\int d^{4}x\Bigl\{\Lambda^{j}\mathcal{G}_{j}-(1+\beta^{2})\omega_{t}^{0j}\epsilon_{jk}{}^{l}K_{b}^{k}E_{l}^{b}\Bigr\}.
\end{aligned}
\end{equation}
Here the covariant derivative with respect to the Ashtekar-Barbero
connection is defined as $\mathcal{D}_{a}^{(A)}X_{i}^{a}=\partial_{a}X_{i}^{a}+\epsilon_{ij}{}^{k}A_{a}^{j}X_{k}^{a}$
and we have introduced the quantity $\Lambda^{i}=\beta\omega_{t}^{0i}-\frac{1}{2}\epsilon_{jk}^{i}\omega_{t}^{jk}$.
%
%
Since neither $\Lambda^{i}$ nor $\omega_{t}{}^{0i}$ appear with
time derivatives in the action, their conjugate momenta must vanish.
Therefore, they serve as Lagrange multipliers of the following constraints
\begin{align}
\mathcal{G}_{i}= & \frac{1}{\kappa\beta}\mathcal{D}_{a}^{(A)}E_{i}^{a}\approx0, & \mathcal{S}_{i}= & \frac{1}{\kappa\beta}\epsilon_{ij}{}^{k}K_{a}^{j}E_{k}^{a}\approx0.
\end{align}
The ``$\approx$'' symbol here indicates weak equality\footnote{In general, a function $f$ is said to be \textit{weakly equal} to a function $g$, denoted $f\approx g$, if they agree on the subspace defined by the primary constraints.}

To ensure consistency, we must impose the compatibility condition of the densitized triad with the spin connection, which yields the Gauss constraint as 
\begin{equation}
\mathcal{G}_{i}=\frac{1}{\kappa}\epsilon_{ij}{}^{k}K_{a}^{j}E_{k}^{a}\,.
\end{equation}
As a result, the constraint $\mathcal{S}_{i}$ becomes proportional
to $\mathcal{G}_{i}$, indicating that $\mathcal{S}_{i}$ is redundant
and can be consistently discarded.

The remaining two terms of the action yeild the Hamiltonian and the
spatial diffeomorphism constraints, respectively 
\begin{equation}
\mathcal{H}=\frac{1}{2\kappa}\frac{E_{i}^{a}E_{j}^{b}}{\sqrt{\text{det}\,h}}\epsilon^{ij}{}_{k}\Bigl\{\mathcal{F}_{ab}^{k}(A)-(1+\beta^{2})\epsilon^{k}{}_{mn}K_{a}^{m}K_{b}^{n}-2\frac{1+\beta^{2}}{\beta}\mathcal{D}_{[a}^{(\Gamma)}K_{b]}^{k}\Bigr\}\,,
\end{equation}
and 
\begin{equation}
\mathcal{H}_{a}=\frac{1}{\kappa\beta}E_{j}^{b}\mathcal{F}_{ab}^{j}(A)-\frac{(1+\beta^{2})}{\kappa\beta}\epsilon^{j}{}_{kl}E_{j}^{b}K_{a}^{k}K_{b}^{l}\,,
\end{equation}
in which 
\begin{equation}
\mathcal{F}_{ab}^{i}(A)=2\partial_{[a}\Gamma_{b]}^{i}+\epsilon^{i}{}_{jk}\Gamma_{a}^{j}\Gamma_{b}^{k}+2\beta\mathcal{D}_{[a}^{(\Gamma)}K_{b]}^{i}+\beta^{2}\epsilon^{i}{}_{jk}K_{a}^{j}K_{b}^{k}
\end{equation}
is the 2-curvature of the Ashtekar connection and $\mathcal{D}^{(\Gamma)}$
is the covariant derivative with respect to the spin connection. Thus,
the full Hamiltonian for a system in which only the gravitational
degrees of freedom are present takes the form 
\begin{equation}
H=\int d^{3}x\left(-\Lambda^{i}\mathcal{G}_{i}+N\mathcal{H}+N^{a}\mathcal{H}_{a}\right)\,.
\end{equation}
As is well-known, this is the sum of three types of constraints 
\begin{align}
\mathcal{G}_{i}= & \frac{1}{\kappa\beta}\mathcal{D}_{a}^{(A)}E_{i}^{a}=\frac{1}{\kappa}\epsilon_{ij}{}^{k}K_{a}^{j}E_{k}^{a}\,,\\
\mathcal{H}= & \frac{1}{2\kappa}\frac{E_{i}^{a}E_{j}^{b}}{\sqrt{\text{det}\,h}}\epsilon^{ij}{}_{k}\Bigl\{\mathcal{F}_{ab}^{k}(A)-(1+\beta^{2})\epsilon^{k}{}_{mn}K_{a}^{m}K_{b}^{n}\Bigr\}\,,\\
\mathcal{H}_{a} & =\frac{1}{\kappa\beta}E_{j}^{b}\mathcal{F}_{ab}^{j}(A)\,.
\end{align}
It is worth noting that, on the constraint surface, the Gauss constraint
vanishes weakly. This implies that, on this surface, the antisymmetric part of the
extrinsic curvature $K_{ab}$ vanishes, rendering $K_{ab}$ symmetric
in its indices, as expected. 

Proceeding further, additional constraints can be derived by varying
the action with respect to suitable combinations of the non-dynamical
variables. Notably, this procedure yields an explicit expression for
the spin connection $\Gamma_{a}^{i}$ in terms of the triad: 
\begin{equation}
\Gamma_{a}^{i}=\frac{1}{2}\epsilon^{ijk}e_{k}^{b}\left(\partial_{a}e_{bj}-\partial_{b}e_{aj}+e_{a}^{l}e_{j}^{c}\partial_{c}e_{bl}\right)\,.\label{spinconn}
\end{equation}
This equation ensure that the connection $\Gamma_{a}^{i}$ is precisely
the spin connection compatible with the triad.

\section{Fermionic Field}
\label{sec:fermions}

The final goal of the present manuscript is analyzing the CED framework (i.e., the case involving both fermions and photons and their interaction). All electrically charged fermions are described by Dirac spinors. In contrast, Majorana fermions ---being their own antiparticles--- must be electrically neutral. Since CED describes interactions between charged particles mediated by photon exchange, the theory inherently involves Dirac fermions. Consequently, we adopt the four-component representation of fermionic fields \cite{Pal:2011pb, Schwartz:2014sze}. For further details, refer to Appendix \ref{App_fermions}.

First of all, it is necessary to study how the inclusion of fermions alters the familiar vacuum scenario.
This system has been previously studied in the literature \cite{Bojowald:2010qpa, Das:2008r,Thiemann:1998QSDV, Jacobson:1988qta, Mercuri:2006wb, PhysRevD.73.084016, Date:2011rd}. Notably, in \cite{Bojowald:2010qpa} and \cite{Das:2008r}, the Dirac matrices themselves were modified to accommodate the curved spacetime geometry. In contrast, the present work adopts an approach in which instead of altering the matrices, we modify the algebraic identities involving them (see Appendix \ref{App_fermions}). This results in a formulation that more mathematically closely mirrors the flat spacetime treatment of fermionic fields, allowing a more straightforward comparison between
the two cases. 

To derive the constraints associated to the fermionic contribution we apply the canonical decomposition, $\mathcal{E}_{I}^{a}=e_{I}^{a}+n^{a}n_{I}$, to the Dirac action 
\begin{equation}
S_{F}=\frac{i}{2}\int d^{4}x|e|\Bigl\{ e_{I}^{\mu}\overline{\Psi}\gamma^{I}\mathfrak{D}_{\mu}\Psi-e_{I}^{\mu}\overline{\mathfrak{D}_{\mu}\Psi}\gamma^{I}\Psi\Bigr\}\,.\label{dirac-action}
\end{equation}
where $n_{I}=-\delta_{I}^{0}$ (in the time gauge), as was done in the vacuum case. As a result, the fermionic action becomes
\begin{align}
S_{F}=\frac{i}{2}\int & d^{4}x\sqrt{\text{det}\,h}\Bigl[\overline{\Psi}\gamma^{I}\left(N\mathcal{E}_{I}^{a}-t^{a}n_{i}+N^{a}n_{I}\right)\mathfrak{D}_{a}\Psi\nonumber \\
 & -\overline{\mathfrak{D}_{a}\Psi}\gamma^{I}\left(N\mathcal{E}_{I}^{a}-t^{a}n_{i}+N^{a}n_{I}\right)\Psi\Bigr]\,.
\end{align}
Once again we encounter three types of terms involving $t^{a},\,N$, and $N^{a}$ . The term containing $t^{a}$ leads to 
\begin{equation}
\frac{i}{2}\int d^{4}x\sqrt{\text{det}\,h}\Bigl[\Psi^{\dagger}\partial_{t}\Psi-(\partial_{t}\Psi^{\dagger})\Psi+\omega_{t}^{ij}\Psi^{\dagger}\sigma_{ij}\Psi\Bigr]\,,\label{eq:ta-term-fermion}
\end{equation}
where the first two terms contribute to the symplectic term, while the last one yields the fermionic contribution to the Gauss constraint. As introduced in Section \ref{vacuum}, the Gauss Lagrange multiplier is defined as
\begin{equation}
\Lambda^{i}=\beta\omega_{t}{}^{0i}-\frac{1}{2}\epsilon^{i}{}_{jk}\omega_{t}{}^{jk}\,.
\end{equation}
From this, we can obtain $\omega_{t}^{ij}$ as 
\begin{equation}
\omega_{t}^{ij}=\beta\epsilon^{ij}{}_{k}\omega_{t}{}^{0i}-\epsilon^{ij}{}_{k}\Lambda^{k}\,.
\end{equation}
Replacing this into \eqref{eq:ta-term-fermion} allows us to identify
the fermionic contributions to both the Gauss constraint and the $\mathcal{S}_{i}$
constraints, respectively 
\begin{align}
 & \mathcal{G}_{i}^{F}=\frac{1}{2}\sqrt{\text{det}\,h}J_{i}\,, &  & \mathcal{S}_{i}^{F}=-\frac{1}{2}\beta\sqrt{\text{det}\,h}J_{i}\,,
\end{align}
where the superscript $F$ denotes the fermionic contribution. Here
$J^{I}$ is the \textit{fermionic axial 4-current} defined as $J^{I}=\overline{\Psi}\gamma_{5}\gamma^{I}\Psi$, whose components in terms of left- and right-chiral
spinors and Pauli matrices are 
\begin{align}
J^{i}= & \overline{\Psi}\gamma_{5}\gamma^{i}\Psi=-\Psi_{R}^{\dagger}\sigma^{i}\Psi_{R}-\Psi_{L}^{\dagger}\sigma^{i}\Psi_{L}\,,\\[2ex]
J^{0}= & \overline{\Psi}\gamma_{5}\gamma^{0}\Psi=-\Psi_{R}^{\dagger}\Psi_{R}+\Psi_{L}^{\dagger}\Psi_{L}\,.
\end{align}
The terms in the fermionic action involving the lapse $N$ and the shift vector $N^{a}$ yield, respectively, the fermionic contributions to the Hamiltonian and spatial diffeomorphism constraints as
\begin{align}
\mathcal{H}^{F}= & -\frac{i}{2}E_{i}^{a}\left(\overline{\Psi}\gamma^{i}\mathfrak{D}_{a}\Psi-\overline{\mathfrak{D}_{a}\Psi}\gamma^{i}\Psi\right)+\sqrt{\text{det}\,h}\,m\,\overline{\Psi}\Psi\,,\\
\mathcal{H}_{a}^{F}= & \frac{i}{2}\sqrt{\text{det}\,h}\left(\overline{\Psi}\gamma^{0}\mathfrak{D}_{a}\Psi-\overline{\mathfrak{D}_{a}\Psi}\gamma^{0}\Psi\right)\,.
\end{align}
It is worth noticing that we recover the Weyl fermion case simply by setting $m=0$ in the above.

By minimally coupling the vacuum action (see Section \ref{vacuum})
to the fermionic action, we obtain the total Hamiltonian of the system,
comprising both the vacuum and the fermionic field contributions.
\begin{equation}
H_{\text{G+F}}=\int d^{3}x\left(-\Lambda^{i}\mathcal{G}_{i}^{\text{G+F}}-\omega_{t}^{0i}\mathcal{S}_{i}^{\text{G+F}}+N\mathcal{H}^{\text{G+F}}+N^{a}\mathcal{H}_{a}^{\text{G+F}}\right)\,,
\end{equation}
where 
\begin{align}
\mathcal{G}_{i}^{\text{G+F}}= & \frac{1}{\kappa\beta}\mathcal{D}_{a}^{(A)}E_{i}^{a}+\frac{1}{2}\sqrt{\text{det}\,h}J_{i},\label{gauss_ferm}\\
\mathcal{S}_{i}^{\text{G+F}}= & -\frac{1+\beta^{2}}{\kappa\beta}\epsilon_{ij}{}^{k}K_{a}^{j}E_{k}^{a}-\frac{1}{2}\beta\sqrt{\text{det}\,h}\,J_{i},\label{eq:s-fermion}\\
\mathcal{H}^{\text{G+F}}= & \frac{1}{2\kappa}\frac{E_{i}^{a}E_{j}^{b}}{\sqrt{\text{det}\,h}}\epsilon^{ij}{}_{k}\Bigl\{\mathcal{F}_{ab}^{k}(A)-(1+\beta^{2})\epsilon^{k}{}_{mn}K_{a}^{m}K_{b}^{n}-2\frac{1+\beta^{2}}{\beta}\mathcal{D}_{[a}^{(\Gamma)}K_{b]}^{k}\Bigr\}\nonumber \\
 & -\frac{i}{2}E_{i}^{a}\left(\overline{\Psi}\gamma^{i}\mathfrak{D}_{a}\Psi-\overline{\mathfrak{D}_{a}\Psi}\gamma^{i}\Psi\right) +\sqrt{\text{det}\,h}\,m\,\overline{\Psi}\Psi,\\
\mathcal{H}_{a}^{\text{G+F}}= & \frac{1}{\kappa\beta}\left(E_{j}^{b}\mathcal{F}_{ab}^{j}(A)-(1+\beta^{2})\epsilon^{j}{}_{kl}E_{j}^{b}K_{a}^{k}K_{b}^{l}\right)-\frac{i}{2}\sqrt{\text{det}\,h}\left(\overline{\Psi}\gamma^{0}\mathfrak{D}_{a}\Psi+\overline{\mathfrak{D}_{a}\Psi}\gamma^{0}\Psi\right).\label{diff_ferm}
\end{align}
For the sake of notational simplicity, we will omit the superscript
``G+F'' in the remaining of this section. Unless otherwise stated,
all expressions should be understood as referring to the total system,
including both the vacuum and the fermionic fields.

The presence of the fermionic field modifies some of the fundamental
relations, in particular, the compatibility condition. This can be
seen by varying the combined vacuum and fermionic actions with respect
to the connection 1-form $\omega_{a}^{IJ}$, which yields the following
modified relation 
\begin{equation}
\frac{1}{\kappa}\mathcal{D}_{b}(|e|e_{K}^{[a}e_{L}^{b]})P^{KL}{}_{IJ}-\frac{1}{4}|e|e_{M}^{a}\epsilon^{M}{}_{IJN}J^{N}=0,
\end{equation}
which consequently leads to
\begin{equation}
\mathcal{D}_{b}(|e|e_{K}^{[a}e_{L}^{b]})=\frac{\kappa}{4}\frac{\beta^{2}}{1+\beta^{2}}|e|\left(\epsilon_{KL}{}^{MN}e_{M}^{a}J_{N}-\frac{1}{\beta}\left(e_{K}^{a}J_{L}-e_{L}^{a}J_{K}\right)\right).\label{comp_cond}
\end{equation}
This implies that the system is no longer torsionless and the canonical variables of the theory now carry an implicit dependence on the fermionic field. We emphasize that the present result is stated explicitly, with consistent conventions for both signatures and notation. In contrast, in \cite{Das:2008r} we find some inconsistencies in the sign conventions, and the derivation of the result is not presented in a fully transparent way. Moreover, in \cite{Thiemann:1998QSDV} the Hamiltonian constraint is computed for gravity coupled to fermions; however,it is not entirely clear to us whether this Hamiltonian is derived directly from the action. This may account for the absence of a term in the modification of the Ashtekar connection, as well as in the fermionic contribution to the Hamiltonian constraint. Moreover, as already noted in \cite{Perez:2005pm}, unlike the vacuum case, where the Barbero-Immirzi parameter plays no role in the equations of motion, here the torsion term is proportional to this parameter, effectively governing the strength of the interaction between the gravitational and fermionic fields. As a result, to enforce the compatibility condition in the presence of fermions, the connection 1-form must be modified as 
\begin{equation}
\begin{aligned}\omega_{a}^{IJ}= & \tilde{\omega}_{a}^{IJ}+C_{a}^{IJ}\\
= & \tilde{\omega}_{a}^{IJ}-\frac{\kappa}{4}\frac{\beta^{2}}{1+\beta^{2}}\left(\epsilon^{IJ}{}_{KL}e_{a}^{K}J^{L}-\frac{2}{\beta}e_{a}^{[I}J^{J]}\right)\,,
\end{aligned}
\end{equation}
where $C_{a}^{IJ}$ is the contorsion tensor and $\tilde{\omega}_{a}^{IJ}$
denotes the torsion-free (vacuum) connection 1-form.


Now recall that $\Gamma_{a}^{i}=-\epsilon^{i}{}_{jk}\omega_{a}^{jk}$.
This relation implies that any modification to the connection 1-form
necessarily alters the spin connection $\Gamma_{a}^{i}$ from its
vacuum form. To derive the explicit expression for the modified spin
connection in terms of the original (vacuum) spin connection and a
correction term involving the fermionic field, we consider the variation
of the Hamiltonian $H_{\text{G+F}}$ with respect to the spin connection.
This variation yields the symmetric component of the modified spin
connection. The antisymmetric component is then obtained directly
from the compatibility condition \eqref{comp_cond}. Together, these
results uniquely determine the full expression for the modified spin
connection as 
\begin{align}
\Gamma_{a}^{i}= & \tilde{\Gamma}_{a}^{i}+C_{a}^{i}\nonumber \\
= & \frac{1}{2}\epsilon^{ijk}e_{k}^{b}\left(\partial_{a}e_{bj}-\partial_{b}e_{aj}+e_{a}^{l}e_{j}^{c}\partial_{c}e_{bl}\right)+\frac{\kappa}{4}\frac{\beta^{2}}{1+\beta^{2}}\left(e_{a}^{i}J^{0}-\frac{1}{\beta}\epsilon^{i}{}_{jk}e_{a}^{j}J^{k}\right),\label{newspin}
\end{align}
where $\Tilde{\Gamma}_{a}^{i}$ is the spin connection in the vacuum
case, and $C_{a}^{i}$ is the spatial projection of the contorsion
tensor $C_{a}^{IJ}$. Hereafter, all the torsionless (i.e.,
the vacuum case) variables will be marked with a tilde, while the
variables with implicit fermionic contribution will be written without. 

Although the modification to the extrinsic curvature can also be analyzed,
it is not immediately relevant for the present discussion and will
be addressed at a later stage in the paper. For now, we observe that
the redefinition of the spin connection leads to a modified Ashtekar-Barbero
connection. Its explicit form, incorporating the fermionic contribution,
is given by 
\begin{equation}
\begin{aligned}A_{a}^{i} & =\beta K_{a}^{i}+\Gamma_{a}^{i}\\
 & =\beta K_{a}^{i}+\tilde{\Gamma}_{a}^{i}+\frac{\kappa}{4}\frac{\beta^{2}}{1+\beta^{2}}\left(e_{a}^{i}J^{0}-\frac{1}{\beta}\epsilon^{i}{}_{jk}e_{a}^{j}J^{k}\right)\,.
\end{aligned}
\label{ash_F}
\end{equation}
With the modified Ashtekar-Barbero connection at hand, we are now
in a position to express the constraints of the theory, explicitly
separating the vacuum contributions from those arising due to the
fermionic sector.

Taking into account the above considerations, the modified compatibility
condition takes the form 
\begin{equation}
\partial_{a}E_{i}^{a}+\epsilon_{ij}{}^{k}\Gamma_{a}^{j}E_{k}^{a}-\frac{\kappa}{4}\frac{\beta^{2}}{1+\beta^{2}}\epsilon_{ij}{}^{k}\left(e_{a}^{j}J^{0}-\frac{1}{\beta}\epsilon^{j}{}_{mn}e_{a}^{m}J^{n}\right)E_{k}^{a}=0\,.
\end{equation}
This condition leads to the following Gauss constraint 
\begin{equation}
\mathcal{G}_{i}=\frac{1}{\kappa}\epsilon_{ij}{}^{k}K_{a}^{j}E_{a}^{k}-\frac{1}{2}\frac{\beta^{2}}{1+\beta^{2}}\sqrt{\text{det}\,h}J_{i}\,
\end{equation}
Once again, we see that the $\mathcal{S}_{i}$ constraint \eqref{eq:s-fermion}
is proportional to the Gauss constraint and can be neglected. Finally,
the first class constraints (detailed calculations in Appendix \ref{App_first_class}) in the case in which gravity is minimally
coupled to a fermionic field are 
\begin{align}
\mathcal{G}_{i}= & \frac{1}{\kappa}\epsilon_{ij}{}^{k}K_{a}^{j}E_{a}^{k}+\frac{1}{2}\frac{\beta^{2}}{1+\beta^{2}}\sqrt{\text{det}\,h}J_{i}\,, \label{g_ferm}\\
\mathcal{H}= & \frac{1}{2\kappa}\frac{E_{i}^{a}E_{j}^{b}}{\sqrt{\text{det}\,h}}\epsilon^{ij}{}_{k}\left(\mathcal{F}_{ab}^{k}(A)-(1+\beta^{2})\epsilon^{k}{}_{mn}K_{a}^{m}K_{b}^{n}-2\frac{1+\beta^{2}}{\beta}\mathcal{D}_{[a}^{(\Gamma)}K_{b]}^{k}\right)\nonumber \\
 & -\frac{i}{2}E_{i}^{a}\left(\overline{\Psi}\gamma^{i}\mathfrak{D}_{a}\Psi-\overline{\mathfrak{D}_{a}\Psi}\gamma^{i}\Psi\right) +\sqrt{\text{det}\,h}\,m\,\overline{\Psi}\Psi\,,\\
\mathcal{H}_{a}= & \frac{1}{\beta\kappa}E_{j}^{b}\left(\mathcal{F}_{ab}^{j}(A)-(1+\beta^{2})\epsilon_{kl}^{j}K_{a}^{k}K_{b}^{l}\right)+\frac{i}{2}\left(\overline{\Psi}\gamma^{0}\mathfrak{D}_{a}\Psi-\overline{\mathfrak{D}_{a}\Psi}\gamma^{0}\Psi\right)\,, \label{ha_ferm}
\end{align}
where $\mathcal{F}_{ab}^{i}(A)$ is the 2-curvature of the modified
Ashtekar connection \eqref{ash_F} and $\mathcal{D}_{a}^{(\Gamma)}$
is the covariant derivative with respect to the new spin connection
of gravity plus fermions. 

Another modification arises in the definition of the fermionic covariant
derivatives, which can now be re-expressed in terms of the covariant
derivative associated with the modified Ashtekar connection, along
with an additional correction term 
\begin{equation}
\begin{aligned}\mathfrak{D}_{a}\Psi & =\partial_{a}\Psi+\frac{1}{2}\omega_{a}{}^{IJ}\sigma_{IJ}\Psi\\
 & =\partial_{a}\Psi+\omega_{a}{}^{0i}\sigma_{0i}\Psi+\frac{1}{2}\omega_{a}{}^{ij}\sigma_{ij}\Psi\\
 & =\partial_{a}\Psi+K_{a}^{i}\sigma_{0i}\Psi-\frac{1}{2}\epsilon^{ij}{}_{k}\Gamma_{a}^{k}\sigma_{ij}\Psi\\
 & =D_{a}^{(A)}\Psi+K_{a}^{i}(1+i\beta\gamma_{5})\sigma_{0i}\Psi\,,
\end{aligned}
\label{fermion-derivative}
\end{equation}
where we have defined the modified covariant derivative 
\begin{equation}
D_{a}^{(A)}\Psi=\partial_{a}\Psi-iA_{a}^{i}\gamma_{5}\sigma_{0i}\Psi\,.\label{D^(A)-fermions}
\end{equation}
As a side remark, we note that there appears to be some inconsistency in the literature regarding the definition of the fermionic covariant derivative. In particular, in \cite{Das:2008r} the same definitions for the spin connection $\Gamma_a^i$ and the Pauli basis $\tau^i$ as those adopted here are used. However, when expressing the covariant derivative \eqref{D^(A)-fermions} in the Pauli basis, the authors obtain $D_a = \partial_a + A_a^i \tau_i$, which is not consistent with the stated definition of the covariant derivative, nor with the resulting expressions for the constraints, which otherwise agree with ours. A similar issue appears in \cite{Thiemann:1998QSDV}. In that case, however, the explicit definition of $\Gamma_a^i$ is not provided, preventing a direct comparison. We note that alternative sign conventions for the spin connection are sometimes adopted in the literature. For completeness, we have verified whether our results remain unchanged under such a modification. We find that not only do the resulting constraints differ from those obtained here, but also additional inconsistencies would arise already at the level of the vacuum constraints (e.g., the vacuum Gauss constraint does not take its broadly accepted form), which are generally assumed to be well established. Our understanding is that this discrepancy may originate from an ambiguity in index conventions, allowing for sign changes without immediate consequences in intermediate expressions. Since the explicit form of the fermionic covariant derivative in the Pauli basis is not required in the remainder of this work, we do not pursue this issue further and leave a more detailed investigation to future studies.
Continuing the computations, for the conjugate fermionic field we find 
\begin{equation}
\overline{\mathfrak{D}_{a}\Psi}=\overline{D_{a}^{(A)}\Psi}+K_{a}^{i}\Psi^{\dagger}(1-i\beta\gamma_{5})\sigma_{0i}\gamma^{0}\label{antifermion-derivative}
\end{equation}
Inserting these modified expressions into the Hamiltonian and diffeomorphism
constraints allows us to compute the new forms of the constraints,
respectively
\begin{align}
\mathcal{H}= & \frac{1}{2\kappa}\frac{E_{i}^{a}E_{j}^{b}}{\sqrt{\text{det}\,h}}\epsilon^{ij}{}_{k}\left(\mathcal{F}_{ab}^{k}(A)-(1+\beta^{2})\epsilon^{k}{}_{mn}K_{a}^{m}K_{b}^{n}-2\frac{1+\beta^{2}}{\beta}\mathcal{D}_{[a}^{(\Gamma)}K_{b]}^{k}\right)\nonumber \\
 & -\frac{i}{2}E_{i}^{a}\left(\overline{\Psi}\gamma^{i}D_{a}^{(A)}\Psi-\overline{D_{a}^{(A)}\Psi}\right){+}\frac{1}{2}\epsilon^{i}{}_{jk}E_{i}^{a}K_{a}^{j}J^{k}+\frac{\beta}{2}E_{i}^{a}K_{a}^{i}J^{0} +\sqrt{\text{det}\,h}\,m\,\overline{\Psi}\Psi\,,
\end{align}
and 
\begin{equation}
\begin{aligned}\mathcal{H}_{a}= & \frac{1}{\beta\kappa}E_{j}^{b}\left(\mathcal{F}_{ab}^{j}(A)-(1+\beta^{2})\epsilon^{j}{}_{kl}K_{a}^{k}K_{b}^{l}\right)\\
 & +\frac{i}{2}\left(\overline{\Psi}\gamma^{0}D_{a}^{(A)}\Psi-\overline{D_{a}^{(A)}\Psi}\gamma^{0}\Psi\right)-\frac{\beta}{2}\sqrt{\text{det}\,h}K_{a}^{i}J_{i}\,.
\end{aligned}
\end{equation}
In the above formulation, the canonical pair for the fermionic system
is $(\Psi,\Pi)$ where $\Pi=i\sqrt{\text{det}\,h}\Psi$. However,
this leads to a symplectic term in the action in which there is an
additional term implying an imaginary correction to the new Ashtekar
connection. This will ultimately lead to a modification of the Poisson
brackets. This issue can be resolved by taking into account the half-density
fermionic fields \cite{Bojowald:2010qpa,Thiemann:1998QSDV, Thiemann:1997rq}. Thus, the new spinors and their conjugate
momenta become 
\begin{align}
\xi= & \sqrt[4]{\text{det}\,h}\Psi, & \pi= & i\xi^{\dagger}.
\end{align}
As a result, we get the correct symplectic term and consequently the
following anti-Poisson brackets for the half-density fermionic fields
and their momenta 
\begin{equation}
\{\xi_{A}(x),\pi_{B}(y)\}_{+}=\delta_{AB}\delta(x-y),
\end{equation}
where $x,\,y$ are points residing on the same spatial hypersurface.
Additionally, the fermionic axial 4-current can also be redefined
by introducing the densitized axial 4-current 
\begin{equation}
\overline{J}^{I}=\sqrt{\text{det}\,h}J^{I}.
\end{equation}
Its components can be expressed in terms of the half-density spinor
fields and the generators $\tau^{i}$ of the $su(2)$ Lie algebra,
namely: 
\begin{align}
\overline{J}^{i}= & \overline{\xi}\gamma_{5}\gamma^{i}\xi=-i\pi\gamma^{0}\gamma_{5}\gamma^{i}\xi=-2\left(\pi_{L}\tau^{i}\xi_{L}+\pi_{R}\tau^{i}\xi_{R}\right)\,,\\
\overline{J}^{0}= & \overline{\xi}\gamma_{5}\gamma^{0}\xi=-i\pi\gamma^{0}\gamma_{5}\gamma^{0}\xi=i(\pi_{R}\xi_{R}-\pi_{L}\xi_{L})\,.
\end{align}
Moreover, one can show that \cite{Thiemann:1998QSDV} 
\begin{equation}
D_{a}^{(A)}\Psi=\frac{1}{\sqrt[4]{\text{det}\,h}}D_{a}^{(A)}\xi\,.\label{eq:DAxi}
\end{equation}
%
%
We are now in a position to verify whether the equations of motion (EOM) derived from this modified theory are consistent with the usual curved spacetime case. To the best of our knowledge, the EOM for this system have not been previously analyzed, and performing this consistency check is essential to ensure the validity of the underlying theory. 

To perform this check, our strategy is to show that the EOM coming
from our Hamiltonian formulation match those directly coming from
the Euler-Lagrange equations. In curved spacetime, where the covariant
derivative is the $D_{a}^{(A)}$ defined in Eq. \eqref{D^(A)-fermions},
the fermionic EOM for $\Psi$ derived from the Euler-Lagrange equation
become 
\begin{equation}
ie_{0}^{a}\gamma^{0}D_{a}^{(A)}\Psi-\frac{1}{2}\beta e_{0}^{a}K_{a}^{i}\gamma_{5}\gamma_{i}\Psi+ie_{i}^{a}\gamma^{i}D_{a}^{(A)}\Psi-\frac{1}{2}\epsilon^{i}{}_{jk}e_{i}^{a}K_{a}^{j}\gamma_{5}\gamma^{k}\Psi-\frac{1}{2}\beta e_{i}^{a}K_{a}^{i}\gamma_{5}\gamma^{0}\Psi=0\,,
\end{equation}
which reduces to the well-known EOM for fermionic field in flat spacetime,
namely the Dirac equation. Replacing $\Psi$ with its expression
in terms of the half-density spinor field $\xi$, and taking into
account the spacetime foliation adopted from the beginning, the above
EOM becomes 
\begin{equation}
(t^{a}-N^{a})[i\gamma^{0}D_{a}^{(A)}\xi-\frac{1}{2}\beta K_{a}^{i}\gamma_{5}\gamma_{i}\xi]-N[ie_{i}^{a}\gamma^{i}D_{a}^{(A)}\xi-\frac{1}{2}\epsilon^{i}{}_{jk}e_{i}^{a}K_{a}^{j}\gamma_{5}\gamma^{k}\xi-\frac{1}{2}\beta e_{i}^{a}K_{a}^{i}\gamma_{5}\gamma_{0}\xi]=0\,.\label{eq:EoM-Lagrange}
\end{equation}
On the other hand, the EOM coming from the Hamiltonia formulation
can be derived using the anti-Poisson brackets as
\begin{equation}
\dot{\xi}=\left\{ \xi(x), H_{\text{G+F}}\right\}
\end{equation}
where the Hamiltonian and diffeomorphism constraints in the above
expression written in terms of half-density fields have the following
forms 
\begin{equation}
\begin{aligned}\mathcal{H}= & \frac{1}{2\kappa}\frac{E_{i}^{a}E_{j}^{b}}{\sqrt{\text{det}\,h}}\epsilon^{ij}{}_{k}\left(\mathcal{F}_{ab}^{k}(A)-(1+\beta^{2})\epsilon^{k}{}_{mn}K_{a}^{m}K_{b}^{n}-2\frac{1+\beta^{2}}{\beta}\mathcal{D}_{[a}^{(\Gamma)}K_{b]}^{k}\right)\\
 & -\frac{1}{2}e_{i}^{a}\left(\pi\gamma^{0}\gamma^{i}D_{a}^{(A)}\xi-D_{a}^{(A)}\pi\gamma^{0}\gamma^{i}\xi\right)-\frac{i}{2}\epsilon^{i}{}_{jk}e_{i}^{a}K_{a}^{j}\pi\gamma^{0}\gamma^{k}\xi-\frac{i\beta}{2}e_{i}^{a}K_{a}^{i}\pi\gamma^{0}\gamma_{5}\gamma^{0}\xi\,,\\
\mathcal{H}_{a}= & \frac{1}{\kappa\beta}E_{j}^{b}\left(\mathcal{F}_{ab}^{j}(A)-(1+\beta^{2})\epsilon_{kl}^{j}K_{a}^{k}K_{b}^{l}\right)+\frac{1}{2}\left(\pi D_{a}^{(A)}\xi-D_{a}^{(A)}\pi\xi\right)+\frac{i\beta}{2}K_{a}^{i}\pi\gamma^{0}\gamma_{5}\gamma^{i}\xi\,.
\end{aligned}
\end{equation}
A straightforward computation shows that by keeping the dependence
of $A_{a}^{i}$ and $\Gamma_{a}^{i}$ on the fermionic field implicit,
these constraints lead to the exact equations of motion as \eqref{eq:EoM-Lagrange}.

Finally, we can simplify the constraints by in terms of the new Ashtekar connection, the half-density fermionic fields and the Pauli basis. This yields:
\begin{align}
\mathcal{G}_{i} & =\frac{1}{\kappa}\epsilon_{ij}{}^{k}K_{a}^{j}E_{k}^{a}-\frac{i}{2}\frac{\beta^{2}}{1+\beta^{2}}\pi\gamma^{0}\gamma_{5}\gamma^{i}\xi\,,\label{Gmass}\\
\mathcal{H} & =\frac{1}{2\kappa}\frac{E_{i}^{a}E_{j}^{b}}{\sqrt{\text{det}\,h}}\epsilon^{ij}{}_{k}\left(\mathcal{F}_{ab}^{k}(A)-(1+\beta^{2})\epsilon^{k}{}_{mn}K_{a}^{m}K_{b}^{n}\right)+\frac{1+\beta^{2}}{\beta^{2}}\mathcal{D}_{a}^{(\Gamma)}(e_{i}^{a}\mathcal{G}^{i})\label{Hmass}\\
 & +\frac{i}{2}\beta e_{i}^{a}D_{a}^{(A)}(\pi\gamma^{0}\gamma_{5}\gamma^{i}\xi)-\frac{1}{8}({3}+2\beta^{2})\frac{\kappa\beta^{2}}{1+\beta^{2}}\frac{1}{\sqrt{\text{det}\,h}}(\pi\gamma^{0}\gamma_{5}\gamma_{i}\xi)(\pi\gamma^{0}\gamma_{5}\gamma^{i}\xi)\nonumber \\
 & -\frac{i\kappa}{4}({-3}+2\beta^{2})\mathcal{G}_{i}(\pi\gamma^{0}\gamma_{5}\gamma^{i}\xi)-\frac{1}{2}e_{i}^{a}\left(\pi\gamma^{0}\gamma^{i}D_{a}^{(A)}\xi-D_{a}^{(A)}\pi\gamma^{0}\gamma^{i}\xi\right)-im\pi\gamma^{0}\,,\\
\mathcal{H}_{a} & =\frac{1}{\kappa\beta}E_{j}^{b}\mathcal{F}_{ab}^{j}(A)-\frac{1+\beta^{2}}{\beta}\mathcal{G}_{i}K_{a}^{i}+\frac{1}{2}\left(\pi D_{a}^{(A)}\xi-D_{a}^{(A)}\pi\xi\right)\,.\label{Hamass}
\end{align}
In contrast to the massive case, when treating with Weyl fermions, it is convenient to use the bispinor notation and express the constraints in terms of the Pauli basis. i.e.
\begin{align}
\mathcal{G}_{i} & =\frac{1}{\kappa}\epsilon_{ij}{}^{k}K_{a}^{j}E_{k}^{a}-\frac{\beta^{2}}{1+\beta^{2}}\left(\pi_{L}\tau_{i}\xi_{L}+\pi_{R}\tau_{i}\xi_{R}\right)\,,\label{G}\\
\mathcal{H} & =\frac{1}{2\kappa}\frac{E_{i}^{a}E_{j}^{b}}{\sqrt{\text{det}\,h}}\epsilon^{ij}{}_{k}\left(\mathcal{F}_{ab}^{k}(A)-(1+\beta^{2})\epsilon^{k}{}_{mn}K_{a}^{m}K_{b}^{n}\right)+\frac{1+\beta^{2}}{\beta^{2}}\mathcal{D}_{a}^{(\Gamma)}(e_{i}^{a}\mathcal{G}^{i})\label{H}\\
 & +\frac{1}{2}(3+2\beta^{2})\frac{\kappa\beta^{2}}{1+\beta^{2}}\frac{1}{\sqrt{\text{det}\,h}}\left(\pi_{L}\tau_{i}\xi_{L}+\pi_{R}\tau_{i}\xi_{R}\right)\left(\pi_{L}\tau^{i}\xi_{L}+\pi_{R}\tau^{i}\xi_{R}\right)\nonumber \\
 & +\beta e_{i}^{a}D_{a}^{(A)}\left(\pi_{L}\tau^{i}\xi_{L}+\pi_{R}\tau^{i}\xi_{R}\right)-\frac{\kappa}{2}(-3+2\beta^{2})\mathcal{G}_{i}\left(\pi_{L}\tau^{i}\xi_{L}+\pi_{R}\tau^{i}\xi_{R}\right)\nonumber \\
 & -ie_{i}^{a}\left(\pi_{R}D_{a}^{(A)}\xi_{R}-\pi_{L}D_{a}^{(A)}\xi_{L}-D_{a}^{(A)}\pi_{R}\xi_{R}+D_{a}^{(A)}\pi_{L}\xi_{L}\right)\,,\\
\mathcal{H}_{a} & =\frac{1}{\kappa\beta}E_{j}^{b}\mathcal{F}_{ab}^{j}(A)-\frac{1+\beta^{2}}{\beta}\mathcal{G}_{i}K_{a}^{i}+\frac{1}{2}\left(\pi_{R}D_{a}^{(A)}\xi_{R}+\pi_{L}D_{a}^{(A)}\xi_{L}-D_{a}^{(A)}\pi_{R}\xi_{R}-D_{a}^{(A)}\pi_{L}\xi_{L}\right)\,.\label{Ha}
\end{align}
From this form of the constraints, we can easily recognize the vacuum contributions and the added fermionic ones, where now all the variables depend either implicitly or explicitly on the fermionic field. When fermionic fields are absent, the standard vacuum constraints are readily recovered.

\section{Canonical Electrodynamics}
\subsection{The full Hamiltonian\label{sec:QED}}
In this section, we examine a system in which photons ---denoted
by a $\gamma$ with no indices--- are coupled to a fermionic field,
and both fields are minimally coupled to gravity. Unlike the classical
formulation of canonical electrodynamics on a fixed flat spacetime, here the
background is allowed to be dynamical. There have been several attempts to couple the gravitational field to electromagnetism \cite{Das:2008r,Bojowald:2010qpa}. However, to the best of our knowledge, no work has considered a framework in which gravity, fermions, and photons are all simultaneously coupled. In addition, in \cite{Das:2008r} a specific choice for the shift vector is adopted, which differs from the standard decomposition. As a consequence, certain time-derivative terms do not cancel in the intermediate steps of the derivation, leading to an extra contribution in the electromagnetic sector of the Hamiltonian.\\
To maintain clarity and avoid
an overload of calculations at once, we begin by analyzing the CED
action in curved spacetime using modified Ashtekar-Barbero variable. After deriving the CED contributions to
the constraints, we will then proceed to minimally couple the full
system to the gravitational field.

The CED action in curved spacetime takes the form: 
\begin{equation}
\begin{aligned}S_{\text{CED}}= & S_{F}+S_{\gamma}\\
= & \int d^{4}x|e|\Bigl\{\frac{i}{2}\left(e_{I}^{a}\overline{\Psi}\gamma^{I}\tilde{\mathfrak{D}}_{a}\Psi-e_{I}^{a}\overline{\mathfrak{D}_{a}\Psi}\gamma^{I}\Psi\right)-m\overline{\Psi}\Psi\Bigr\}\\
 & +\int d^{4}\sqrt{-\text{det}\,g}\left(-\frac{1}{4}g^{ac}g^{bd}F_{cd}F_{ab}\right)\,.
\end{aligned}
\label{QEDaction}
\end{equation}
where the fermionic covariant derivatives contain an interaction term
between the photon and the fermionic field, as follows 
\begin{align}
\tilde{\mathfrak{D}}_{a}\Psi= & \partial_{a}\Psi+\frac{1}{2}\omega_{a}{}^{IJ}\sigma_{IJ}\Psi+i\,q\,A_{a}\Psi\,,\\
\overline{\tilde{\mathfrak{D}}_{a}\Psi}= & \partial_{a}\overline{\Psi}+\frac{1}{2}\omega_{a}{}^{IJ}\Psi^{\dagger}\sigma_{IJ}^{\dagger}\gamma^{0}-i\,q\,A_{a}\overline{\Psi}\,.
\end{align}
In Eq. \eqref{QEDaction} and in the following, $A_{a}$ denotes the
electromagnetic 4-vector potential, $q$ the electric charge and $F_{ab}=(\nabla_{a}A_{b}-\nabla_{b}A_{a})$
is the electromagnetic field tensor with $\nabla_{a}$ being the covariant
derivative with respect to the Christoffel symbol.

Let us begin by focusing solely on the photonic contribution, described
by the action $S_{\gamma}$. Given that the spacetime metric decomposes
as $g^{ab}=h^{ab}-n^{a}n^{b}$, its contraction with the EM field
tensors becomes 
\begin{equation}
\begin{aligned}g^{ac}g^{bd}F_{cd}F_{ab}= & h^{ac}h^{bd}F_{cd}F_{ab}-\frac{2}{N^{2}}h^{bd}(t^{a}-N^{a})(\nabla_{a}A_{b}-\nabla_{b}A_{a})(t^{c}-N^{c})\\
 & \times(\nabla_{c}A_{d}-\nabla_{d}A_{c})\\
= & h^{ac}h^{bd}F_{cd}F_{ab}-\frac{2}{N^{2}}h^{bd}\left(\dot{A}_{b}-\nabla_{b}(t^{a}A_{a})-N^{a}F_{ab}\right)\\
 & \times\left(\dot{A}_{d}-\nabla_{d}(t^{c}A_{c})-N^{c}F_{cd}\right)\,.
\end{aligned}
\label{dec_Fab}
\end{equation}
Substituting this expression into the photonic action yields: 
\begin{equation}
\begin{split}S_{\gamma}= & -\frac{1}{4}\int d^{4}x\sqrt{\text{det}\,h}\Bigl\{ Nh^{ac}h^{bd}F_{cd}F_{ab}-\frac{2}{N}h^{bd}\left(\dot{A}_{b}-\nabla_{b}(t^{a}A_{a})-N^{a}F_{ab}\right)\\
 & \times\left(\dot{A}_{d}-\nabla_{d}(t^{c}A_{c})-N^{c}F_{cd}\right)\Bigr\}\,.
\end{split}
\end{equation}
Next, we express this action in the Hamiltonian form as $S_{\gamma}=\int d^{4}x\left(\pi^{a}\dot{A}_{a}-H_{\gamma}\right)$,
in which $\pi^{a}$ denotes the momentum conjugate to the spatial
components of the gauge field $A_{a}$. This momentum is computed
via functional differentiation
\begin{equation}
\begin{aligned}\pi^{a} & =\frac{\delta S_{\gamma}}{\delta\dot{A}_{a}}\\
 & =\frac{\sqrt{\text{det}\,h}}{N}h^{ab}\left(\dot{A}_{b}-\nabla_{b}(t^{c}A_{c})-N^{c}F_{cb}\right)\\
 & =\sqrt{\text{det}\,h}h^{ab}n^{c}F_{cb}\,,
\end{aligned}
\label{pi-A}
\end{equation}
where in the last last we have used part of the second term of \eqref{dec_Fab}.
From
this result, we observe that $\pi^{a}$ is a densitized vector field.
Substituting this expression back into the action, we obtain the action
in the desired form as 
\begin{equation}
\begin{aligned}S_{\gamma}= & \int d^{4}x\Bigl\{-\sqrt{\text{det}\,h}\frac{N}{4}h^{ac}h^{bd}F_{cd}F_{ab}+\frac{1}{2}\frac{N}{\sqrt{\text{det}\,h}}h_{ab}\pi^{a}\pi^{b}\Bigr\}\\
= & \int d^{4}x\bigg\{\pi^{a}\dot{A}_{a}+t^{b}A_{b}\nabla_{a}\pi^{a}-N\left(\frac{\sqrt{\text{det}\,h}}{4}h^{ac}h^{bd}F_{cd}F_{ab}+\frac{1}{2\sqrt{\text{det}\,h}}h_{ab}\pi^{a}\pi^{b}\right)\\
 & -N^{b}F_{ba}\pi^{a}\bigg\}\,.
\end{aligned}
\end{equation}
Once again one finds four terms corresponding to symplectic term,
and Gauss, Hamiltonian and spatial diffeomorphism constraints. In
particular, since the temporal component of the gauge field $A_{a}$
does not possess a conjugate momentum, it acts as a Lagrange multiplier.
Specifically, $t^{b}A_{b}=A_{t}$ enforces a constraint, which leads
to the presence of the Gauss constraint associated with $U(1)$ gauge
symmetry. The full set of constraints for the electromagnetic field
in the Hamiltonian formulation becomes 
\begin{align}
\mathcal{G}^{\gamma} & =\nabla_{a}\pi^{a}\,,\\
\mathcal{H}^{\gamma} & =\frac{\sqrt{\text{det}\,h}}{4}h^{ac}h^{bd}F_{cd}F_{ab}+\frac{1}{2\sqrt{\text{det}\,h}}h_{ab}\pi^{a}\pi^{b}\,,\\
\mathcal{H}_{a}^{\gamma} & =F_{ab}\pi^{b}\,.
\end{align}
Let us now perform the same analysis for the full CED action $S_{\text{CED}}$
in \eqref{QEDaction}. Beginning with the contributions involving
the time vector $t^{a}$, we identify the following terms
\begin{equation}
\begin{split}\int d^{4}x & \Bigl[\pi^{a}\dot{A}_{a}+\frac{i}{2}\sqrt{\text{det}\,h}\left(\Psi^{\dagger}\partial_{t}\Psi-(\partial_{t}\Psi)\Psi\right)\Bigr]\\
 & +\int d^{4}x\left\{ \frac{i}{2}\sqrt{\text{det}\,h}\,\omega_{t}{}^{ij}\Psi^{\dagger}\sigma_{ij}\Psi+A_{t}\left(\nabla_{a}\pi^{a}-\sqrt{\text{det}\,h}\,q\Psi^{\dagger}\Psi\right)\right\} \,.
\end{split}
\end{equation}
We observe that the first integral corresponds to the symplectic term,
while the second integral contains two Gauss constraints, each multiplied
by a Lagrange multiplier, respectively $\omega_{t}^{ij}$ and $A_{t}$.
One of these constraints depends solely on internal indices and represents
the fermionic contribution to the vacuum Gauss constraint. The other
involves only spacetime indices ---specifically, the spatial ones---
and reduces to Gauss' Law in the absence of the fermionic field. Hence,
the full CED contribution to the Gauss constraint is 
\begin{equation}
\Lambda\mathcal{G}^{\text{CED}}=\omega_{t}{}^{ij}\,\frac{i}{2}\sqrt{\text{det}\,h}\,\Psi^{\dagger}\sigma_{ij}\Psi+A_{t}\left(\nabla_{a}\pi^{a}-\sqrt{\text{det}\,h}\,q\Psi^{\dagger}\Psi\right)\,.
\end{equation}
As in the previous section, the Lagrange multiplier $\omega_{t}^{ij}$
will be expressed in terms of $\Lambda_{i}$ and $\omega_{t}^{0i}$
in order to derive its contribution to the vacuum Gauss and $\mathcal{S}_{i}$
constraints.

Next, considering the terms involving the lapse function $N$, we
obtain the CED Hamiltonian constraint as
\begin{equation}
\begin{split}\mathcal{H}^{\text{CED}} & =\frac{\sqrt{\text{det}\,h}}{4}h^{ac}h^{bd}F_{cd}F_{ab}+\frac{1}{2\sqrt{\text{det}\,h}}h_{ab}\pi^{a}\pi^{b}-\frac{i}{2}E_{i}^{a}\left(\overline{\Psi}\gamma^{i}\tilde{\mathfrak{D}}_{a}\Psi-\overline{\tilde{\mathfrak{D}}_{a}\Psi}\gamma^{i}\Psi\right)\\
 & +\sqrt{\text{det}\,h}\,m\overline{\Psi}\Psi\,.
\end{split}
\end{equation}
Finally, the terms multiplied by the shift vector $N^{a}$ yield the
CED diffeomorphism constraint 
\begin{equation}
\mathcal{H}_{a}^{\text{CED}}=F_{ab}\pi^{b}+\frac{i}{2}\sqrt{\text{det}\,h}\left(\overline{\Psi}\gamma^{0}\tilde{\mathfrak{D}}_{a}\Psi-\overline{\tilde{\mathfrak{D}}_{a}\Psi}\gamma^{0}\Psi\right)\,.
\end{equation}
It is now possible to minimally couple the gravitational field to
the photonic and fermionic fields in order to obtain the full constraints
of the system, which are 
\begin{align}
\Lambda^{i}\mathcal{G}_{i}^{\text{G+F}}+A_{t}\mathcal{G}^{\text{CED}}= & \left(\beta\omega_{t}{}^{0i}-\frac{1}{2}\epsilon^{i}{}_{jk}\omega_{t}{}^{jk}\right)\Bigl[\frac{1}{\kappa\beta}\left(\partial_{a}E_{i}^{a}+\epsilon_{ip}{}^{q}A_{a}^{p}E_{q}^{a}\right)+\frac{1}{2}\sqrt{\text{det}\,h}J_{i}\Bigr]\nonumber \\
 & +A_{t}\left(\nabla_{a}\pi^{a}-\sqrt{\text{det}\,h}q\Psi^{\dagger}\Psi\right)\,,\label{gaussQED}\\
\mathcal{S}_{i}^{\text{Full}}= & -\frac{1+\beta^{2}}{\kappa\beta}\epsilon_{ij}{}^{k}K_{a}^{j}E_{k}^{a}-\frac{1}{2}\beta\sqrt{\text{det}\,h}J_{i}\,,\nonumber \\
\mathcal{H}^{\text{Full}}= & \frac{1}{2\kappa}\frac{E_{i}^{a}E_{j}^{b}}{\sqrt{\text{det}\,h}}\epsilon^{ij}{}_{k}\left(\mathcal{F}_{ab}^{k}(A)-(1+\beta^{2})\epsilon^{k}{}_{mn}K_{a}^{m}K_{b}^{n}-2\frac{1+\beta^{2}}{\beta}\mathcal{D}_{[a}^{(\Gamma)}K_{b]}^{k}\right)\nonumber\\
 & +\frac{\sqrt{\text{det}\,h}}{4}h^{ac}h^{bd}F_{cd}F_{ab}+\frac{1}{2\sqrt{\text{det}\,h}}h_{ab}\pi^{a}\pi^{b}\nonumber \\
 &-\frac{i}{2}E_{i}^{a}\left(\overline{\Psi}\gamma^{i}\tilde{\mathfrak{D}}_{a}\Psi-\overline{\tilde{\mathfrak{D}}_{a}\Psi}\gamma^{i}\Psi\right) +\sqrt{\text{det}\,h}m\overline{\Psi}\Psi\,,\\
\mathcal{H}_{a}^{\text{Full}}= & \frac{1}{\kappa\beta}E_{i}^{b}\left(\mathcal{F}_{ab}^{i}(A)-(1+\beta^{2})\epsilon^{i}{}_{jk}K_{a}^{j}K_{b}^{k}\right)\nonumber \\
&+\frac{i}{2}\sqrt{\text{det}\,h}\left(\overline{\Psi}\gamma^{0}\tilde{\mathfrak{D}}_{a}\Psi-\overline{\tilde{\mathfrak{D}}_{a}\Psi}\gamma^{0}\Psi\right)
  +F_{ab}\pi^{b}\,.\label{diffQED}
\end{align}
None of the terms involving the 4-vector potential $A_{a}$ couples
to the Lorentz connection 1-form $\omega_{a}^{IJ}$. As a result,
varying the full action, where the gravitational field is coupled
to both fermionic and photonic fields, with respect to the 1-form
yields the same result as in the case where only gravity and the fermionic
field are present, 
\begin{equation}
\frac{\delta S_{\text{Full}}}{\delta\omega_{a}{}^{IJ}}=\frac{\delta S_{F}}{\delta\omega_{a}{}^{IJ}}=-\frac{1}{4}|e|e_{K}^{a}\epsilon^{K}{}_{IJL}J^{L}\,.
\end{equation}
Thus, as expected, photons do not imply torsion. Hence, the compatibility
condition remains the following 
\begin{equation}
\mathcal{D}_{a}\left(|e|e^{[c}{}_{K}e^{a]}{}_{L}\right)=\frac{\kappa}{4}\frac{\beta^{2}}{1+\beta^{2}}|e|\Bigl[\epsilon_{KL}{}^{MN}e_{M}^{c}J_{N}-\frac{1}{\beta}\left(e_{K}^{c}J_{L}-e_{L}^{C}J_{K}\right)\Bigr]\,.
\end{equation}
Consequently, the new spin connection takes the same form as in Eq.\eqref{newspin}.
Moreover, the full constraints in terms of the modified Ashtekar connection
look like Eqs. \eqref{gaussQED}-\eqref{diffQED}, but now $A_{a}^{i}$
implicitly depends on the fermionic field given in Eq. \eqref{ash_F}.

Following the steps we performed before, we can now reformulate the
fermionic covariant derivatives in terms of the modified Ashtekar
connection, making its dependence on this connection explicit. They
are written as follows 
\begin{align}
\tilde{\mathfrak{D}}_{a}\Psi= & D_{a}^{(A)}\Psi+K_{a}^{i}\left(1+i\beta\gamma_{5}\right)\sigma_{0i}\Psi+iqA_{a}\Psi\,,\\
\overline{\tilde{\mathfrak{D}}_{a}\Psi}= & \overline{D_{a}^{(A)}\Psi}+K_{a}^{i}\Psi^{\dagger}\left(1-i\beta\gamma_{5}\right)\sigma_{0i}\gamma^{0}-iqA_{a}\overline{\Psi}\,,
\end{align}
where $D_{a}^{(A)}$ has been defined in Eq. \eqref{D^(A)-fermions}.
Hence, the Hamiltonian constraint become 
\begin{align}
\mathcal{H}^{\text{Full}}= & \frac{1}{2\kappa}\frac{E_{i}^{a}E_{j}^{b}}{\sqrt{\text{det}\,h}}\epsilon^{ij}{}_{k}\left(\mathcal{F}_{ab}^{k}(A)-(1+\beta^{2})\epsilon^{k}{}_{mn}K_{a}^{m}K_{b}^{n}-2\frac{1+\beta^{2}}{\beta}\mathcal{D}_{[a}^{(\Gamma)}K_{b]}^{k}\right)\nonumber \\
 & +\frac{\sqrt{\text{det}\,h}}{4}h^{ac}h^{bd}F_{cd}F_{ab}+\frac{1}{2\sqrt{\text{det}\,h}}h_{ab}\pi^{a}\pi^{b}-\frac{i}{2}E_{i}^{a}\left(\overline{\Psi}\gamma^{i}D_{a}^{(A)}\Psi-\overline{D_{a}^{(A)}\Psi}\gamma^{i}\Psi\right)\nonumber \\
 & {+}\frac{1}{2}\epsilon^{i}{}_{jk}E_{i}^{a}K_{a}^{j}J^{k}+\frac{\beta}{2}E_{i}^{a}K_{a}^{i}J^{0}+qE_{i}^{a}A_{a}\mathcal{J}^{i}+\sqrt{\text{det}\,h}\,m\overline{\Psi}\Psi\,,
\end{align}
where $\mathcal{J}^{I}=\overline{\Psi}\gamma^{I}\Psi$ is the fermionic
4-current, and the spatial diffeomorphism constraint turns out to
be
\begin{align}
\mathcal{H}_{a}^{\text{Full}}= & \frac{1}{\kappa\beta}E_{i}^{b}\left(\mathcal{F}_{ab}^{i}(A)-(1+\beta^{2})\epsilon^{i}{}_{jk}K_{a}^{j}K_{b}^{k}\right)+\frac{i}{2}\sqrt{\text{det}\,h}\left(\overline{\Psi}\gamma^{0}D_{a}^{(A)}\Psi-\overline{D_{a}^{(A)}\Psi}\gamma^{0}\Psi\right)\nonumber \\
 & -\frac{\beta}{2}\sqrt{\text{det}\,h}K_{a}^{i}J_{i}-\sqrt{\text{det}\,h}qA_{a}\mathcal{J}^{0}+F_{ab}\pi^{b}\,.
\end{align}
Next, we turn our attention to the EOM. First, we derive the EOM using
the Euler-Lagrange equation. The fermionic covariant derivative remains
the same as in the Dirac fermions section. However, an additional
term now appears, which is the electromagnetic coupling.
Consequently, in curved spacetime, the EOM for the fermionic field
$\Psi$ takes the form
\begin{equation}
\begin{split}ie_{0}^{a}\gamma^{0}D_{a}^{(A)}\Psi & -\frac{1}{2}\beta e_{0}^{a}K_{a}^{i}\gamma_{5}\gamma_{i}\Psi+ie_{i}^{a}\gamma^{i}D_{a}^{(A)}\Psi{-}\frac{1}{2}\epsilon^{i}{}_{jk}e_{i}^{a}K_{a}^{j}\gamma_{5}\gamma^{k}\Psi\\
 & -\frac{1}{2}\beta e_{i}^{a}K_{a}^{i}\gamma_{5}\gamma^{0}\Psi-m\Psi-qA_{a}e_{0}^{a}\gamma^{0}\Psi-qA_{a}e_{i}^{a}\gamma^{i}\Psi=0\,,
\end{split}
\end{equation}
which is consistent with the well-known CED EOM for fermions. On the
other hand, the EOM for the photon field can be obtained by extremizing
the full action $S_{\text{Full}}$ with respect to the electromagnetic
4-vector potential $A_{a}$. This yields a background-independent
form of Maxwell's equations. More concretely, the photonic Euler-Lagrange
equation reads 
\begin{equation}
\frac{\delta\mathcal{L}^{\text{CED}}}{\delta A_{b}}-\nabla_{c}\frac{\delta\mathcal{L}^{\text{CED}}}{\delta\left(\nabla_{c}A_{b}\right)}=0\,,
\end{equation}
which leads to the following expression for the photonic EOM in curved
spacetime 
\begin{equation}
\nabla_{c}\left(g^{ac}g^{bd}F_{ad}\right)=qe_{I}^{b}\mathcal{J}^{I}\,,\label{photonEOM}
\end{equation}
which is consistent with the CED theory.

We can rewrite the fermionic EOM in terms of the half-density fermionic
field and the foliation of the spacetime introduced from the beginning,
obtaining 
\begin{align}
(t^{a}-N^{a}) & \Bigl[i\gamma^{0}D_{a}^{(A)}\xi-\frac{\beta}{2}K_{a}^{i}\gamma_{5}\gamma_{i}\xi-qA_{a}\gamma^{0}\xi\Bigr]-N\Bigl[ie_{i}^{a}\gamma^{i}D_{a}^{(A)}\xi{-}\frac{1}{2}\epsilon^{i}{}_{jk}e_{i}^{a}K_{a}^{j}\gamma_{5}\gamma^{k}\xi\nonumber \\
 & -\frac{\beta}{2}e_{i}^{a}K_{a}^{i}\gamma_{5}\gamma^{0}\xi-m\xi-qA_{a}e_{i}^{a}\gamma^{i}\xi\Bigr]=0\,.\label{eq:QED-Psi-EOM-EL}
\end{align}
On the other hand, in the Hamiltonian formulation, we can substitute
the half-density field into the Hamiltonian and diffeomorphism constraints,
and then use the fermionic anti-Poisson brackets, to derive the Hamiltonian
EOM. Doing so, we recover the same result as obtained through the
Euler-Lagrange method in \eqref{eq:QED-Psi-EOM-EL}. 

As for the photonic field, additional manipulation of the equation
derived from the Euler-Lagrange method is required before it can be
directly compared to the result obtained from the Hamiltonian formalism.
By substituting $g^{ab}=h^{ab}-n^{a}n^{b}$ and expressing $\pi^{a}$
in terms of the electromagnetic field tensor $F_{ab}$ (Eq. \eqref{pi-A})
in the Euler-Lagrange EOM, the photonic field EOM becomes
\begin{equation}
\nabla_{c}\left(h^{ac}h^{bd}F_{ad}\right)+\nabla_{c}\left(\frac{1}{\sqrt{\text{det}\,h}}n^{b}\pi^{c}\right)-\nabla_{c}\left(\frac{1}{\sqrt{\text{det}\,h}}n^{c}\pi^{b}\right)-qe_{I}^{b}\mathcal{J}^{I}=0\,.
\end{equation}
This is the EOM for the conjugate momentum $\pi^{a}$ of the 4-vector
potential $A_{a}$. 

On the other hand, in order to obtain the EOM via the Hamiltonian
formalism, we need to consider the following photonic Poisson bracket
\begin{equation}
\{A_{a}(x),\pi^{b}(y)\}=\delta_{a}^{b}\delta(x-y)\,,
\end{equation}
which implies 
\begin{equation}
\dot{\pi}^{a}=-\frac{\delta H^{\text{Full}}}{\delta A_{a}}\,.
\end{equation}
Substituting the full Hamiltonian in the previous equation, we obtain
the expected result, consistent with the Euler-Lagrange method. 

Finally, once again we impose the compatibility condition and observe
that the $\mathcal{S}_{i}$ constraint is proportional to the Gauss
constraint arising from the fermionic modification of the vacuum.
As a result, $\mathcal{S}_{i}$ can be discarded. Therefore, the remaining
(first class) constraints are the Gauss constraint derived in the
previous section and the one associated with the CED sector, together
with the Hamiltonian and the diffeomorphism constraints. Moreover,
since we are working with Dirac spinors ---and more generally within
the framework of CED--- the considerations regarding chirality remain
applicable. Consequently, the final full constraints, expressed in
terms of the modified Ashtekar-Barbero connection and the half-density
fermionic field, are 
\begin{align}
\Lambda^{i}\mathcal{G}_{i}^{\text{G+F}}+A_{t}\mathcal{G}^{\text{CED}}= & \Lambda^{i}\Bigl[\frac{1}{\kappa}\epsilon_{ij}{}^{k}K_{a}^{j}E_{k}^{a}-\frac{i}{2}\frac{\beta^{2}}{1+\beta^{2}}\pi\gamma^{0}\gamma_{5}\gamma^{i}\xi\Bigr]+A_{t}\left(\nabla_{a}\pi^{a}+iq\pi\xi\right)\,,\\
\mathcal{H}^{\text{Full}}= & \frac{1}{2\kappa}\frac{E_{i}^{a}E_{j}^{b}}{\sqrt{\text{det}\,h}}\epsilon^{ij}{}_{k}\left(\mathcal{F}_{ab}^{k}(A)-(1+\beta^{2})\epsilon^{k}{}_{mn}K_{a}^{m}K_{b}^{n}\right)\nonumber \\
 & +\frac{1+\beta^{2}}{\beta^{2}}\mathcal{D}_{a}^{(\Gamma)}\left(e_{i}^{a}\mathcal{G}^{i}\right)+\frac{i}{2}\beta e_{i}^{a}D_{a}^{(A)}(\pi\gamma^{0}\gamma_{5}\gamma^{i}\xi)\nonumber \\
 & -\frac{1}{8}({3}+2\beta^{2})\frac{\kappa\beta^{2}}{1+\beta^{2}}\frac{1}{\sqrt{\text{det}\,h}}(\pi\gamma^{0}\gamma_{5}\gamma_{i}\xi)(\pi\gamma^{0}\gamma_{5}\gamma^{i}\xi)\nonumber \\
 & -\frac{i\kappa}{4}({-3}+2\beta^{2})\mathcal{G}_{i}(\pi\gamma^{0}\gamma_{5}\gamma^{i}\xi)-im\pi\gamma^{0}\xi\nonumber \\
 & -\frac{1}{2}e_{i}^{a}\left(\pi\gamma^{0}\gamma^{i}D_{a}^{(A)}\xi-D_{a}^{(A)}\pi\gamma^{0}\gamma^{i}\xi\right)-iqe_{i}^{a}A_{a}\pi\gamma^{0}\gamma^{i}\xi\nonumber \\
 & +\frac{\sqrt{\text{det}\,h}}{4}h^{ac}h^{bd}F_{cd}F_{ab}+\frac{1}{\sqrt{\text{det}\,h}}h_{ab}\pi^{a}\pi^{b}\,,\\
\mathcal{H}_{a}^{\text{Full}}= & \frac{1}{\kappa\beta}E_{j}^{b}\mathcal{F}_{ab}^{j}(A)-\frac{1+\beta^{2}}{\beta}\mathcal{G}_{i}K_{a}^{i}+\frac{1}{2}\left(\pi D_{a}^{(A)}\xi-D_{a}^{(A)}\pi\xi\right)\nonumber \\
 & +iqA_{a}\pi\xi+F_{ab}\pi^{b}\,.
\end{align}

\subsection{Parity transformation\label{sec:parity}}

In this final subsection, we analyze the behavior of the full system
under parity transformation. A similar study was conducted in \cite{Das:2008r}
for a system where gravity is coupled solely to a fermionic field.
Here, we extend the analysis to the full system, where the gravitational
field is minimally coupled to both fermionic and photonic fields,
providing clearer, more detailed, and more readable final results.

To determine the transformation properties of the new Ashtekar connection
under parity, we first need to identify the torsion contribution to
the extrinsic curvature. We decompose the extrinsic curvature $K_{a}^{i}$
into a torsionless part $\tilde{K}_{a}^{i}$ and a torsion contribution
$k_{a}^{i}$, as follows
\begin{equation}
K_{a}^{i}=\tilde{K}_{a}^{i}+k_{a}^{i}\,.
\end{equation}
In order to provide and explicit expression for $k_{a}^{i}$, the
starting point is the EOM of the canonical variables $(A_{a}^{i},E_{i}^{a})$,
which are 
\begin{align}
\mathcal{L}_{t}A_{a}^{i}= & \left\{ A_{a}^{i},H^{\text{\text{Full}}}\right\} =\kappa\beta\frac{\delta H^{\text{\text{Full}}}}{\delta E_{i}^{a}}\,,\\
\mathcal{L}_{t}E_{i}^{a}= & \left\{ E_{i}^{a},H^{\text{\text{Full}}}\right\} =-\kappa\beta\frac{\delta H^{\text{\text{Full}}}}{\delta A_{a}^{i}}\,,\label{EOMtriad}
\end{align}
where$\mathcal{L}_{t}$ denotes the Lie derivative with respect to the time function $t$ and 
the full Hamiltonian
$H^{\text{\text{Full}}}=\int d^{3}x\left(-\Lambda\mathcal{G}^{\text{\text{Full}}}+N\mathcal{H}^{\text{\text{Full}}}+N^{a}\mathcal{H}_{a}^{\text{\text{Full}}}\right)$
refers to the system in which gravity is minimally coupled to both
photon and fermionic fields which also interact among themselves.
The first equation governs the dynamics of the system, while the second
provides the expression for the connection. Therefore, we focus only
on Eq. \eqref{EOMtriad}. By varying the full Hamiltonian with respect
to the modified Ashtekar connection, we obtain
\begin{align}
\mathcal{L}_{t}E_{p}^{c}= & \left(\frac{1}{2}\epsilon^{i}{}_{jk}\omega_{t}^{jk}-\beta\omega_{t}^{0i}\right)\epsilon_{ip}{}^{n}E_{n}^{c}\nonumber \\
 & +N\left[\frac{1}{\beta\sqrt{\text{det}\,h}}\left(E_{p}^{c}E_{m}^{a}A_{a}^{m}-E_{m}^{c}E_{p}^{a}A_{a}^{m}\right)-\frac{1}{\beta}\partial_{a}\left(\epsilon^{acb}e_{bp}\right){+}\frac{1}{\kappa}\epsilon^{i}{}_{kp}E_{i}^{c}J^{k}\right]\nonumber \\
 & +N^{a}\partial_{a}E_{p}^{c}+E_{p}^{c}\partial_{a}N^{a}-E_{p}^{b}\partial_{b}N^{c}-\epsilon^{j}{}_{kp}N^{a}E_{j}^{c}A_{a}^{k}+\frac{\kappa}{\beta}N^{c}\mathcal{G}_{p}^{\text{G+F}}
\end{align}
It is worth noting that, as in the vacuum case,
the parity of the extrinsic curvature ---and, consequently, that
of the entire system--- can be defined only on the constraint hypersurface.
In particular, we must impose the vanishing of the Gauss constraint
$\mathcal{G}^{\text{G+F}}$, which will allow us to express the extrinsic
curvature in terms of the fermionic axial four-current. However, we
will later observe that the same dependence of the extrinsic curvature
on the fermionic field could also be derived directly from the modified
connection 1-form obtained via the compatibility condition. Nevertheless,
this alternative route does not resolve the issue of determining the
behavior of the system under parity transformations. Even in that
case, one must still restrict to the hypersurface defined by $\mathcal{G}^{\text{G+F}}=0$
in order to properly analyze the parity properties of the variables
of the system. Hence, we neglect the last term in the previous expression.
Contracting the previous equation with $e_{cq}$ and $\delta^{pq}$,
and then symmetrizing the result, we obtain 
\begin{equation}
\begin{split}e_{cq}\mathcal{L}_{t}E_{p}^{c} & +e_{cp}\mathcal{L}_{t}E_{q}^{c}+N\frac{1}{\beta}\epsilon^{acb}\left(e_{cq}\partial_{a}e_{bp}+e_{cp}\partial_{a}e_{bq}-\delta_{pq}e_{c}^{i}\partial_{a}e_{bi}\right)\\
 & +\frac{N}{\beta}\sqrt{\text{det}\,h}\left(e_{p}^{c}A_{cq}+e_{q}^{c}A_{cp}\right)+\sqrt{\text{det}\,h}\partial_{b}N^{c}\left(e_{cq}e_{p}^{b}+e_{cp}e_{q}^{b}\right)=0\,.
\end{split}
\label{eq:isol-K}
\end{equation}
To isolate the extrinsic curvature, we first expand the term involving
the Ashtekar connection as 
\begin{equation}
\begin{split}\left(e_{p}^{c}A_{cq}+e_{q}^{c}A_{cp}\right)= & \beta\left(e_{p}^{c}\tilde{K}_{cq}+e_{q}^{c}\tilde{K}_{cp}\right)+\beta\left(e_{p}^{c}k_{cq}+e_{q}^{c}k_{cp}\right)+\left(e_{p}^{c}\tilde{\Gamma}_{cq}+e_{q}^{c}\tilde{\Gamma}_{cp}\right)\\
 & +\left(e_{p}^{c}C_{cq}+e_{q}^{c}C_{cp}\right)\,.
\end{split}
\end{equation}
Then, substituting this expression into Eq. \eqref{eq:isol-K} and
using the definition of the torsionless spin connection in terms of
the triad \eqref{spinconn} and the torsion contribution to the spin
connection, we find 
\begin{equation}
\begin{split}e_{cq}\mathcal{L}_{t}E_{p}^{c} & +e_{cp}\mathcal{L}_{t}E_{q}^{c}+\sqrt{\text{det}\,h}\partial_{b}N^{c}\left(e_{cq}e_{p}^{b}+e_{cp}e_{q}^{b}\right)\\
 & +\frac{N}{\beta}\sqrt{\text{det}\,h}\left[\beta\left(e_{p}^{c}\tilde{K}_{cq}+e_{q}^{c}\tilde{K}_{cp}\right)+\beta\left(e_{p}^{c}k_{cq}+e_{q}^{c}k_{cp}\right)+\frac{\kappa}{2}\frac{\beta^{2}}{1+\beta^{2}}\delta_{pq}J^{0}\right]=0\,.
\end{split}
\label{eq:Lpia}
\end{equation}
To isolate the torsionless part of the extrinsic curvature, we use
its definition 
\begin{equation}
\tilde{K}_{ab}=\frac{1}{2N}\left(\dot{h}_{ab}-2\tilde{D}_{(a}N_{b)}\right)\,,
\end{equation}
where $\tilde{D}_{a}N_{b}=h_{a}^{c}h_{b}^{d}\nabla_{c}N_{d}$ with
$\nabla_{a}$ the covariant derivative with respect to the Christoffel
symbols. Contracting this expression with the triad and the densitized
triad and symmetrizing yields 
\begin{equation}
E_{p}^{c}\tilde{K}_{cq}+E_{q}^{c}\tilde{K}_{cp}=-\frac{1}{N}\left(e_{cq}\mathcal{L}_{t}E_{p}^{c}+e_{cp}\mathcal{L}_{t}E_{q}^{c}+(E_{p}^{c}e_{q}^{d}+E_{q}^{c}e_{p}^{d})\partial_{c}N_{d}\right)\,.
\end{equation}
Substituting this back into Eq.\eqref{eq:Lpia}, the symmetric torsion
contribution to the extrinsic curvature is found to obey 
\begin{equation}
e_{p}^{c}k_{cq}+e_{q}^{c}k_{cp}=-\frac{1}{2}\frac{\kappa\beta}{1+\beta^{2}}\delta_{pq}J^{0}\,.
\end{equation}
For the antisymmetric expression, we can take advantage of the vanishing
Gauss constraint $\mathcal{G}^{\text{G+F}}$ and separate the torsionless
and torsion contribution to the extrinsic curvature as
\begin{equation}
\mathcal{G}_{i}^{\text{G+F}}=\frac{1}{\kappa}\epsilon_{ij}{}^{k}(\tilde{K}_{a}^{j}+k_{a}^{j})E_{k}^{a}+\frac{1}{2}\frac{\beta^{2}}{1+\beta^{2}}\sqrt{\text{det}\,h}J_{i}=0\,,
\end{equation}
Since we are considering the constraint hypersurface, the torsionless
extrinsic curvature is symmetric. Hence, the antisymmetric part of
the torsion contribution to the extrinsic obeys
\begin{equation}
e_{q}^{c}k_{cp}-e_{p}^{c}k_{cq}=-\frac{1}{2}\frac{\kappa\beta^{2}}{1+\beta^{2}}\epsilon^{i}{}_{pq}J_{i}\,.
\end{equation}
By summing the symmetric and antisymmetric components, we derive the
expression for the torsion contribution to the extrinsic curvature
in terms of the fermionic axial 4-current as 
\begin{equation}
k_{a}^{i}=-\frac{\kappa}{4}\frac{\beta^{2}}{1+\beta^{2}}\left(\frac{1}{\beta}e_{a}^{i}J^{0}+\epsilon^{i}{}_{jk}e_{a}^{j}J^{k}\right)\,.
\end{equation}
As a consequence, the complete splitting of the Ashtekar connection
into torsionless and torsion contributions becomes
\begin{equation}
\begin{split}A_{a}^{i} & =\left(\beta\tilde{K}_{a}^{i}+\tilde{\Gamma}_{a}^{i}\right)+\left(\beta k_{a}^{i}+C_{a}^{i}\right)\\
 & =\left(\beta\tilde{K}_{a}^{i}+\tilde{\Gamma}_{a}^{i}\right)-\frac{\kappa\beta}{4}\epsilon^{i}{}_{jk}e_{a}^{j}J^{k}\,.
\end{split}
\end{equation}
where $C_{a}^{i}$ denotes the spatial projection of the contorsion tensor as defined in \eqref{newspin}. In the second line, we have used the explicit forms of $C_{a}^{i}$ and $k_{a}^{i}$ to express their contribution in terms of the fermionic current.

We can now examine the behavior of the various variables under
a parity transformation. It is important to specify whether we are
referring to internal or spacetime indices, as the Gauss constraint
can be decomposed according to the two distinct Lagrange multipliers,
one of which involves only spacetime indices. Under a parity transformation,
the components associated with the internal indices of the triad change
orientation while the fermionic field transform as
\begin{align}
\mathds{P}e_{i}^{a}= & -e_{i}^{a}\,,\\
\mathds{P}\Psi= & \gamma^{0}\Psi\,,
\end{align}
where $\mathds{P}$ is the parity operator. Moreover, the components
of the fermionic axial 4-current and of the fermionic 4-current transform
respectively as
\begin{align}
\mathds{P}J^{0}= & -J^{0}\,, & \mathds{P}J^{i}= & J^{i}\,,
\end{align}
and
\begin{align}
\mathds{P}\mathcal{J}^{0}= & \mathcal{J}^{0}\,, & \mathds{P}\mathcal{J}^{i}= & -\mathcal{J}^{i}\,.
\end{align}
The torsionless part of the spin connection is proportional to an
even number of triads, while the torsion contribution is proportional
to the triad and different components of the fermionic axial 4-current.
Thus, the spin connection transforms under parity as 
\begin{equation}
\mathds{P}\Gamma_{a}^{i}=\tilde{\Gamma}_{a}^{i}+\frac{\kappa}{4}\frac{\beta^{2}}{1+\beta^{2}}\left(e_{a}^{i}J^{0}+\frac{1}{\beta}\epsilon^{i}{}_{jk}e_{a}^{j}J^{k}\right)\,.
\end{equation}
On the other hand, the torsionless extrinsic curvature is contracted
with the triad $\tilde{K}_{a}^{i}=e^{bi}\tilde{K}_{ab}$, hence the
extrinsic curvature transform as 
\begin{equation}
\mathds{P}K_{a}^{i}=-\tilde{K}_{a}^{i}-\frac{\kappa}{4}\frac{\beta^{2}}{1+\beta^{2}}\left(\frac{1}{\beta}e_{a}^{i}J^{0}-\epsilon^{i}{}_{jk}e_{a}^{j}J^{k}\right)\,.
\end{equation}
By combining the transformation rules for triad, spin connection,
extrinsic curvature, and fermionic axial 4-current, we derive the
transformation rule for the Ashtekar-Barbero connection which becomes
\begin{equation}
\mathds{P}A_{a}^{i}=\tilde{\Gamma}_{a}^{i}-\beta\left(\tilde{K}_{a}^{i}-\frac{\kappa}{4}\epsilon^{i}{}_{jk}e_{a}^{j}J^{k}\right)\,.
\end{equation}
For transformation of spacetime indices, we must analyze how the covariant
derivative involving Christoffel symbols behaves under a parity transformation.
To this end, we consider a generic spacetime 4-vector $V^{\mu}$ and
apply the parity transformation operator to it as
\begin{equation}
\mathds{P}V^{\mu}=P_{\nu}^{\mu}V^{\nu},
\end{equation}
where $P_{\nu}^{\mu}$ are the components of the parity operator,
given by
\begin{equation}
P_{\nu}^{\mu}=\begin{pmatrix}1 & 0 & 0 & 0\\
0 & -1 & 0 & 0\\
0 & 0 & -1 & 0\\
0 & 0 & 0 & -1
\end{pmatrix}\,.
\end{equation}
The transfromation of a covector $V_{\mu}$ under parity is then expressed
as 
\begin{equation}
\mathds{P}V_{\mu}=(\left(P^{-1}\right)_{\mu}^{\nu}V_{\nu}.
\end{equation}
Accordingly, the covariant derivative involving the Christoffel connection,
when applied to a 4-covector, transforms under parity as
\begin{align}
\mathds{P}\left(\nabla_{\mu}V_{\nu}\right)= & \left(P^{-1}\right)_{\mu}^{\sigma}\partial_{\sigma}\left[\left(P^{-1}\right)_{\nu}^{\rho}V_{\rho}\right]\nonumber \\
 & -\frac{1}{2}g^{\rho\sigma}\left[\left(P^{-1}\right)_{\mu}^{\tau}\partial_{\tau}g_{\sigma\nu}+\left(P^{-1}\right)_{\nu}^{\tau}\partial_{\tau}g_{\sigma\mu}-\left(P^{-1}\right)_{\sigma}^{\tau}\partial_{\tau}g_{\mu\nu}\right]\,.
\end{align}
Consequently, if $\mu$ is a spatial index, the transformation rule
is $\mathds{P}(\nabla_{a}V_{\nu})=-\nabla_{a}(\mathds{P}V_{\nu})$.
Note that we use $a,\,b,\ldots$ indices as abstract indices (either
spacetime of spatial) but here we meant a spatial abstract index. 

A 4-vector with purely spacetime indices, such as the electromagnetic
4-vector potential $A^{a}$ , transforms under parity as
\begin{align}
\mathds{P}A^{t}= & A^{t}\,, & \mathds{P}A^{a}= & -A^{a}\,,
\end{align}
with the same transformation properties valid for the covector components. With a slight abuse of notation, we use $t$ to denote the time component and $a$ for the spatial components of the spacetime index.
Accordingly, the electromagnetic field strength tensor transforms
as
\begin{align}
\mathds{P}F^{ta}= & =\mathds{P}\left(\nabla^{t}A^{a}-\nabla^{a}A^{t}\right)=-F^{ta}\,,\\
\mathds{P}F^{ab}= & =\mathds{P}\left(\nabla^{a}A^{b}-\nabla^{b}A^{a}\right)=F^{ab}\,.
\end{align}
Finally, the conjugate momentum $\pi^{a}$ associated
with the spatial components of the electromagnetic 4-vector potential
transforms as 
\begin{equation}
\mathds{P}\pi^{a}=-\pi^{a}\,.
\end{equation}
Having established how the various variables transform under parity,
we are able to study the behavior of the symplectic term and the constraints
of the system. The symplectic term can be shown to transforms as
\begin{equation}
\begin{split}\mathds{P}\left(\frac{1}{\kappa\beta}\int d^{4}xE_{i}^{a}\mathcal{L}_{t}A_{a}^{i}\right)= & \frac{1}{\kappa\beta}\int d^{4}x(-E_{i}^{a})\mathcal{L}_{t}\left(\tilde{\Gamma}_{a}^{i}-\beta\left(\tilde{K}_{a}^{i}-\frac{\kappa}{4}\epsilon^{i}{}_{jk}e_{a}^{j}J^{k}\right)\right)\\
= & \frac{1}{\kappa\beta}\int d^{4}xE_{i}^{a}\mathcal{L}_{t}\left(\beta\left(\tilde{K}_{a}^{i}-\frac{\kappa}{4}\epsilon^{i}{}_{jk}e_{a}^{j}J^{k}\right)\right)\\
= & \frac{1}{\kappa\beta}\int d^{4}xE_{i}^{a}\mathcal{L}_{t}A_{a}^{i}\,.
\end{split}
\end{equation}
In deriving this result, we used the fact that $\tilde{\Gamma}_{a}^{i}$
is a function of the triad and, consequently, has a vanishing Poisson
bracket with the densitized triad, i.e., $\{E_{i}^{a},\tilde{\Gamma}_{b}^{j}\}=0$.
This result implies that the symplectic term is invariant under parity
transformation. 

For the Gauss constraint, we do not need to analyse the case in which
the Lagrange multiplier has internal indices, since, as mentioned
before, we are performing our analysis on the hypersurface in which
$\mathcal{G}_{i}^{\text{G+F}}=0$. For 
the terms in the Gauss constraint that depend only on the spacetime
indices, we obtain 
\begin{equation}
\mathcal{G}^{\text{CED}}=\nabla_{a}\pi^{a}-\sqrt{\text{det}\,h}\,q\mathcal{J}^{0}\,.
\end{equation}
Using the transformation rules for spacetime indices, we can conclude
that this part of the constraint is also invariant under parity. 

Applying the parity operator to the diffeomorphism constraint and
considering again the hypersurface $\mathcal{G}_{i}^{\text{G+F}}=0$,
one obtains
\begin{align}
\mathds{P}\left(\mathcal{H}_{a}^{\text{\text{Full}}}\right)= & \mathcal{H}_{a}^{\text{\text{Full}}}-\frac{2}{\kappa\beta}\Bigl[E_{i}^{b}\partial_{a}\tilde{\Gamma}_{b}^{i}+\tilde{\Gamma}_{a}^{i}\partial_{b}E_{i}^{b}+\epsilon^{i}{}_{jk}\tilde{\Gamma}_{a}^{j}\tilde{\Gamma}_{b}^{k}E_{i}^{b}\nonumber \\
 & +\beta^{2}\epsilon^{i}{}_{jk}\left(\tilde{K}_{a}^{j}-\frac{\kappa}{4}\epsilon^{j}{}_{mn}e_{a}^{m}J^{n}\right)\left(\tilde{K}_{b}^{k}-\frac{\kappa}{4}\epsilon^{k}{}_{pq}e_{b}^{p}J^{q}\right)\sqrt{\text{det}\,h}e_{i}^{b}\Bigr]-\sqrt{\text{det}\,h}\beta\tilde{K}_{a}^{i}J_{i}\nonumber \\
= & \mathcal{H}_{a}^{\text{\text{Full}}}-\frac{1}{\kappa\beta}\Bigl[2E_{i}^{b}\partial_{a}\tilde{\Gamma}_{b}^{i}+\beta^{2}\epsilon^{i}{}_{jk}\left(\tilde{K}_{a}^{j}-\frac{\kappa}{4}\epsilon^{j}{}_{mn}e_{a}^{m}J^{n}\right)(-\kappa J_{j}\sqrt{\text{det}\,h})\Bigr]\nonumber \\
 & -\sqrt{\text{det}\,h}\beta\tilde{K}_{a}^{i}J_{i}\nonumber \\
= & \mathcal{H}_{a}^{\text{\text{Full}}}-\frac{2}{\kappa\beta}E_{i}^{b}\partial_{a}\tilde{\Gamma}_{b}^{i}\nonumber \\
= & \mathcal{H}_{a}^{\text{\text{Full}}}-\frac{2}{\kappa\beta}\epsilon_{ij}{}^{k}\tilde{\Gamma}_{b}^{i}\tilde{\Gamma}_{a}^{j}E_{k}^{b}\nonumber \\
= & \mathcal{H}_{a}^{\text{\text{Full}}}\,.
\end{align}
In the above derivation, we have used 
the fact that $\epsilon_{ij}{}^{k}E_{k}^{a}\tilde{\Gamma}_{a}^{i}=0$
with $\tilde{\Gamma}_{a}^{i}$ written in terms of the triad.

Finally, we examine the behavior of the Hamiltonian constraint under
a parity transformation. All previously established comments and transformation
rules remain valid. Therefore, 
after a series of straightforward yet lengthy calculations, we conclude
that the Hamiltonian constraint is also invariant under parity transformation,
i.e.,
\begin{align}
\mathds{P}\mathcal{H}^{\text{\text{Full}}}= & \mathcal{H}^{\text{\text{Full}}}+\frac{1}{2\kappa}\frac{E_{i}^{a}E_{j}^{b}}{\sqrt{\text{det}\,h}}\epsilon^{ij}{}_{k}\left(-4\mathcal{D}_{a}^{(\tilde{\Gamma})}\tilde{K}_{b}^{k}+\beta\kappa\mathcal{D}_{a}^{(\tilde{\Gamma})}(\epsilon^{k}{}_{pq}e_{b}^{p}J^{q})\right)+\beta E_{i}^{a}\mathcal{D}_{a}^{(\tilde{\Gamma})}J^{i}\nonumber \\
= & \mathcal{H}^{\text{\text{Full}}}-\frac{2}{\kappa}\frac{E_{i}^{a}E_{j}^{b}}{\sqrt{\text{det}\,h}}\epsilon^{ij}{}_{k}\mathcal{D}_{a}^{(\tilde{\Gamma})}\tilde{K}_{b}^{k}\nonumber \\
= & \mathcal{H}^{\text{\text{Full}}}
\end{align}
As a consequence of the analysis in this subsection, we conclude that
the theory demonstrates full invariance under parity transformations,
which serves as another consistency check of our formulation.

\section{Discussion and conclusions \label{sec:Conclusion}}

In this work we have provided the full analysis of the canonical formulation of the system of gravity, fermions, and photons, and their interaction in a background-independent way. In particular, we have considered the interaction term of the CED, which to our knowledge, has not been considered in previous studies. 

We have analyzed the constraint structure of the system in the Ashtekar-Barbero variables with fermionic field, and have shown how the connection variables are modified due to the presence of fermions. Furthermore, the expression for the Hamiltonian, spatial diffeomorphism and Guass constraints are given in a way that the contribution of each of the sectors, i.e., gravity, fermions, and photons, are clear.  
We have also derived the EOM of the matter sector, which reduce to the usual EOM of field theory on fixed curved or flat spacetime in the appropriate limits. Finally, the behavior of the system under parity transformation is studied in detail.

This analysis and the results pave the way for further canonical analysis of realistic matter in the universe involved in electromagnetic interactions in strong and dynamical (non-fixed) gravity regimes. Most importantly, this work opens the door for further detailed studies of an interactive system (CED in a dynamical background) using loop quantum gravity, polymer quantization, generalized uncertainty principle, or other theory that uses canonical quantization techniques. A rather immediate application of this analysis is in symmetry-reduced systems such as cosmological models involving electromagnetic matter, or collapse scenarios in spherical symmetric or rotating black holes, which up to now, have been done almost exclusively only with scalar fields.

\acknowledgments{
The authors acknowledge the support of the Natural Sciences and Engineering
Research Council of Canada (NSERC). We also would like to thank Mehdi Assanioussi and Klaus Liegener for insightful discussions.
}


\appendix


\section{Brief complementary information on fermions}
\label{App_fermions}

To remain consistent with standard QFT conventions
\cite{Schwartz:2014sze,Weinberg:1995s,Peskin:1995ev},
we adopt the signature $(+,-,-,-)$ in this appendix only. In the main text, we revert to the GR signature $(-,+,+,+)$.

In curved spacetime, fermions couple to gravity through the spin connection $\omega_\mu^{IJ}$. The spinor covariant derivatives acting on $\Psi$ and its adjoint are defined as
\begin{align}
\mathfrak{D}_\mu\Psi
&=\partial_\mu\Psi+\tfrac{1}{2}\omega_\mu^{IJ}\sigma_{IJ}\Psi,
\label{derferm} \\
\overline{\mathfrak{D}_\mu\Psi}&=\partial_\mu\overline{\Psi}
+\tfrac{1}{2}\omega_\mu^{IJ}\overline{\Psi}\,\sigma_{IJ},
\label{derantiferm}
\end{align}
where 
\begin{equation}
\sigma_{IJ}=\tfrac{1}{4}[\gamma_{I},\gamma_{J}],\,\{\gamma^{I},\gamma^{J}\}=-2\eta^{IJ},\,\gamma^{5}=i\gamma^{0}\gamma^{1}\gamma^{2}\gamma^{3}=\begin{pmatrix}-\mathbb{1} & 0\\
0 & \mathbb{1}
\end{pmatrix}.
\end{equation}
No redefinition of the fermionic fields or gamma matrices is required when passing to curved spacetime; all gravitational effects enter through the tetrad and spin connection.

\section{First class nature of constraints}
\label{App_first_class}
A constraint is said to be \textit{first class} if its Poisson bracket with all other constraints vanishes \textit{weakly}, i.e. on the constraint surface. Essentially if the Poisson bracket leads to a linear combination of other first class constraints. In order to show that the constraints given in Eqs. \eqref{g_ferm}-\eqref{ha_ferm} are first class, we explicitly compute their mutual Poisson brackets.

For the constraints in this paper, one can show this by first evaluating the functional derivatives of each constraint with respect to the canonical variables $A_c^p$ and $E_p^c$,
\begin{align}
    \frac{\delta\mathcal{G}_i}{\delta A_c^p} & =  0 \,,\\
    \frac{\delta\mathcal{G}_i}{\delta E^c_p} &= \frac{1}{\kappa}\varepsilon_{ij}\,^k K_c^j \,,\\
    \frac{\delta\mathcal{H}}{\delta A_c^p} &= \frac{1}{\kappa}\left[\partial_a\left(\frac{E^c_iE^a_j}{\sqrt{\text{det}h}}\varepsilon^{ij}\,_p \right) - \frac{1}{\sqrt{\text{det}h}}\varepsilon^{mn}\,_p(-E^c_m\varepsilon_{nk}\,^lA_a^kE^a_l + E^a_m\varepsilon_{nk}\,^lA_a^kE^c_l) \right] \nonumber\\
    &=  \frac{1}{\sqrt{\text{det}h}}\varepsilon^{mn}\,_p (E^c_m\mathcal{D}_aE^a_n + E^a_n\mathcal{D}_aE^c_m) \approx 0 \,,\\
    \frac{\delta\mathcal{H}}{\delta E^c_p} &=  \frac{1}{\kappa}\frac{1}{\sqrt{\text{det}\,h}}\epsilon^{ij}{}_{k}E^a_i\left(\mathcal{F}_{ac}^{k}(A)-(1+\beta^{2})\epsilon^{k}{}_{mn}K_{a}^{m}K_{c}^{n}-2\frac{1+\beta^{2}}{\beta}\mathcal{D}_{[a}^{(\Gamma)}K_{c]}^{k}\right)\nonumber \\
    & -\frac{i}{2}\left(\overline{\Psi}\gamma^{p}\mathfrak{D}_{c}\Psi-\overline{\mathfrak{D}_{c}\Psi}\gamma^{p}\Psi\right)\,,\\  
    \frac{\delta\mathcal{H}_a}{\delta A_c^p} &= \frac{1}{\beta\kappa} (-\partial_aE^c_p + \delta_a^c \partial_b E_p^b + \varepsilon^j\,_{pl}E^b_jA_b^l\delta_a^c + \varepsilon^j\,_{kp} E_j^cA_a^k)\,,\\
    \frac{\delta\mathcal{H}_a}{\delta E^c_p} &= \frac{1}{\beta\kappa} (\mathcal{F}_{ac}^{p}(A) - (1+\beta^2)\varepsilon^p\,_{kl}K_a^kK_c^l)\,.
\end{align}
In deriving these expressions, we have made repeated use of the compatibility condition $\mathcal{D}_aE^a_i \approx 0 $. Using the above and the canonical Poisson brackets, the algebra of constraints reduce to
\begin{align}
    \{\mathcal{G}_i, \mathcal{H}\} &\approx 0\,,\\
     \{\mathcal{G}_i, \mathcal{H}_a\} &= - \frac{1}{\beta\kappa^2} (-\varepsilon_{ij}\,^pK_c^j\mathcal{D}_aE^c_p + \varepsilon_{ij}\,^pK_a^j\mathcal{D}_bE^b_p) \approx 0\,,\\
      \{\mathcal{H}, \mathcal{H}_a\} &= \frac{1}{\beta\kappa^2}\frac{1}{\sqrt{\text{det}h}} \varepsilon^{ip}\,_k\left(\mathcal{F}_{ac}^{k}(A)-(1+\beta^{2})\epsilon^{k}{}_{mn}K_{a}^{m}K_{c}^{n}-2\frac{1+\beta^{2}}{\beta}\mathcal{D}_{[a}^{(\Gamma)}K_{c]}^{k}\right)\nonumber\\
      &\times\left(-E^a_i\mathcal{D}_aE^c_p + E^c_i\mathcal{D}_bE_p^b\right)  + \frac{i}{2\beta\kappa}\bigl[-\mathcal{D}_aE^c_p\left(\overline{\Psi}\gamma^{p}\mathfrak{D}_{c}\Psi-\overline{\mathfrak{D}_{c}\Psi}\gamma^{p}\Psi\right) \nonumber\\
      &+\mathcal{D}_bE^b_p\left(\overline{\Psi}\gamma^{p}\mathfrak{D}_{a}\Psi-\overline{\mathfrak{D}_{a}\Psi}\gamma^{p}\Psi\right) \bigr] \approx0\,,
\end{align}
which implies that that the Gauss, Hamiltonian, and diffeomorphism constraints close under the Poisson bracket and are all first class.

\bibliographystyle{JHEP}
\bibliography{mainbib}

\end{document}